\documentclass[reprint,
superscriptaddress,
 amsmath,amssymb,
 aps,
prb,
notitlepage,
10pt,
twocolumn
]{revtex4-2}

\usepackage{chemformula}
\usepackage{dcolumn}
\usepackage{graphicx}
\usepackage{amsfonts}
\usepackage{color}
\usepackage{xspace}
\usepackage{bm}
\usepackage{bbold}
\usepackage{xcolor}
\usepackage{float}
\usepackage{epstopdf}
\usepackage{hyperref}
\usepackage[capitalise]{cleveref}

\begin{document}
\title{Exact mapping from short-ranged harmonic elastic models of spin crossover materials to Ising models with  interactions at all length scales}
\author{Nadeem Natt}
\email{nadeemnatt1@gmail.com}
\affiliation{School of Mathematics and Physics, The University of Queensland, QLD 4071, Australia}
\author{Gian Ruzzi}
\affiliation{School of Mathematics and Physics, The University of Queensland, QLD 4071, Australia}
\author{Jace Cruddas}
\affiliation{School of Psychological Sciences, Turner Institute for Brain and Mental Health, Monash Biomedical Imaging, Monash University, Clayton, VIC, Australia}
\author{Ross H. McKenzie}
\affiliation{School of Mathematics and Physics, The University of Queensland, QLD 4071, Australia}
\author{Ben J. Powell}
\affiliation{School of Mathematics and Physics, The University of Queensland, QLD 4071, Australia}
\begin{abstract}
	Spin crossover (SCO) materials are reversible molecular switches found in a wide range of  transition metal complexes and metal organic frameworks (MOFs). They exhibit diverse spin state orderings and transitions between them. Here we present an exact mapping from harmonic elastic models to Ising-like models with both a short-range Ising interaction that decays with a power law at large distances and a long-range (infinite-range) Husimi-Temperley interaction that is independent of distance. We apply this mapping to a simple model of SCO frameworks. 
	This provides a microscopic justification for an Ising-Husimi-Temperley model description, which has previously only been justified on phenomenological grounds.
	Elastic frustration is required for non-zero Ising interactions,  but whether or not the short-range interactions in the  Ising model are geometrically frustrated depends on the ratio of the bulk and shear moduli, or equivalently Poisson's ratio. The long-range interaction has two origins: (i) a self-interaction on the  average spin state, mediated through the coupling between the average spin state and  the unit cell parameters; and (ii) an infrared divergence in the spin state-spin state coupling mediated by displacements of metals and ligands within the unit cell. 
	The long-range interaction is independent of the distance between metal centers, nevertheless it  leads to an extensive contribution to the (free) energy. In this model the Husimi-Temperley interaction dominates transitions of spin states, whereas multistep transitions and intermediate order are observed if only the pure (power-law) Ising interaction are retained, only single-step transitions are found in the full model. 

    
\end{abstract}
\maketitle

\section{Introduction}

Spin crossover molecules can exist in one of two different electronic states, one low spin (LS) and one high spin (HS), with pronounced differences in geometry, magnetism, and colour. 
A transition between these states can be induced by a range of physical perturbations, such as temperature ~\cite{hauser2004}, light  irradiation~\cite{CoboLightSCO2008}, pressure~\cite{Ksenofontov2004}, magnetic fields~\cite{Bousseksou2004}, and electric fields~\cite{PrinsElectricSCO2011}.
The SCO phenomenon has been well documented in octahedral transition metal complexes with $d^4-d^7$ electron configurations~\cite{pavlik2013}.
The molecular bistability of these systems has attracted great interest from the nanoscience community, as this phenomenon has potential applications in data storage and display devices~\cite{Gutlich2004, pavlik2013}. This interest can be evidenced by the fact that since the discovery of thermally induced SCO, several hundreds of SCO complexes have been synthesised and studied~\cite{Bousseksou2011}. Furthermore, the change in electronic configuration is accompanied by drastic structural (volume, shape)~\cite{Guionneau2014}, colour~\cite{Gutlich2004} and magnetic changes; these characteristics could have potential applications in mechanical nanoscale machines~\cite{Mullaney2017}, smart pigments, and optical switches~\cite{Guionneau2014}.

SCO systems have been reported to present collective phenomena including hysteresis, multistep transitions, gradual transitions,
and a variety of spin-state orderings, such as anti-ferroelastic phases~\cite{pavlik2013, Bousseksou2011}. Cooperative interactions between molecules are a key factor in understanding spin-state phase transitions; as a consequence, experimental advances have been accompanied by theoretical investigations aimed at determining the physical origin of this cooperativity. A wide variety of models have been proposed to explain the microscopic origin of spin-state transitions; Ising-like models, specially the Wajnflasz and Pick (WP) model~\cite{wajnflasz}, macroscopic thermodynamical models (such as that due to Slichter and Drickamer \cite{Slichter}), and Landau type models \cite{Collet-landau}, have been extensively used because of their simplicity and generality~\cite{pavlik2013}. However, the origin of their interaction parameters is unclear given the dramatic simplifications they employ. Refs.~\onlinecite{pavlik2013, Bousseksou2011}  review a range of models that have been used. All of these models reproduce some characteristics, like gradual and abrupt spin-state transitions, multi-step transitions, incomplete transitions, and hysteretic behaviour~\cite{pavlik2013,Bousseksou2011,Watanabe2016}.

Even though the previously mentioned models have been able to capture many observed phenomena, recent advances in experimental methods,
particularly in determining the spatial ordering of spin states, underline the necessity of more realistic theoretical approaches~\cite{Bousseksou2011}. It is believed that the cooperativity between molecules has an elastic origin; this interaction arises from the pronounced volume change (\textit{ca.} 25\% in FeN$_6$ environments~\cite{Guionneau2014}) of the SCO molecules upon spin change~\cite{Bousseksou2011}. This difference in volume leads to elastic strains and local deformations of the system's lattice, which may in turn lead to complicated long- and short-range interactions. This suggests that constructing effective models should lead to the emergence of both short- and long-range interactions. 

It is unfortunate that the terms short- and long-range have been defined in many different ways by different authors (see \cite{Defenu} for a review of this literature). To avoid the potential confusion arising from this, in the discussion below (which is limited to a model with local classical binary degrees of freedom, i.e., pseudospins) we will refer to any interaction that vanishes as the separation between pseudospins goes to infinite 
as `Ising interactions' and interactions that are independent of the seperation between pseudospins as  Husimi-Temperley interactions. Interactions that depend on the spatial separation of the pseudospins but do not vanish as the separation goes to infinite 
can trivially be rewritten as the sum of an Ising and an Husimi-Temperley interaction and therefore will be referred to as Ising-Husimi-Temperley interactions. We endeavor to limit the use of ``short-range" and ``long-range" to qualitative statements.

Models that explicitly take into account structural degrees of freedom have been proposed, the majority of them are ball and spring models with a lattice mismatch. They have been solved via a variety of methods including purely Monte Carlo methods~\cite{KonishiMC2008,Traiche2018},
purely molecular dynamics methods~\cite{nishino2007,NishinoElastic2009}, and a combination of Monte Carlo methods with damped dynamics~\cite{EnachescuMC2012,Paez2016}. These models exhibit
first-order and gradual one- and two-step phase transitions.
Models have also been proposed that treat the quantum mechanical transitions between spin states microscopically as well as the elastic interactions between molecules \cite{NadeemJACS,NadeemCS,Shuang,NadeemNano}. 
It was observed some time ago that the structure of the SCO crystals qualitatively influences the observed spin-crossover phenomena \cite{HalcrowStructure, HalcrowHysteresis}; however, only recently have detailed structure-property relationships begun to be identified  \cite{JaceSquare}.

Here we present a method, based on a displaced oscillator transformation, to exactly map an elastic system of balls and springs to an Ising-Husimi-Temperley model. We apply this mapping to the square lattice, but the method can readily be applied to other geometries and other elastic models. We show that the Ising coupling constants arise purely from the elastic interactions (not the dynamics of the system) and give explicit relationships between the bulk and shear moduli and the Ising parameters. The Husimi-Temperley interactions do not decay with distance, but nevertheless lead to an extensive contribution to the (free) energy, as required physically \cite{Defenu}. They arise from two independent sources, the coupling of the average spin state to the average unit cell size and a infrared discontinuity in the interactions mediated by changes in the positions of metals/ligands within the unit cell.
We also note the relevance of our model to the physically similar problem of elastic interactions in binary alloys \cite{FRATZL,Frechette1,frechette2021elastic,wagner1974elastic}.


\section{Mapping an elastic model to an Ising model}\label{sec:SquareDerivation}

\subsection{Elastic model}

\begin{figure}
	\begin{center}
	\includegraphics[width=0.95\columnwidth]{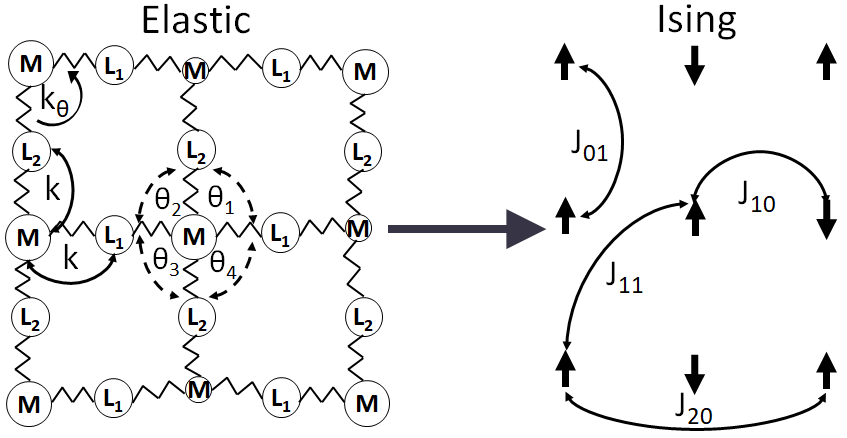}
	\caption{Mapping from a balls and springs model to an Ising model. (Left) Schematic illustration of a square lattice of metal ions (M) connected to ligands (L$_1$, L$_2$) by springs; smaller circles represent metals in the LS state, and larger circles in the HS state; $k$ is the elastic constant of the springs, and $k_\theta$ corresponds to the elastic constant for  bending. Each $M$ has four corresponding angle variables $\theta$, see Eq. (\ref{eq:HSquareElastics}). (Right) Schematic illustration of the corresponding Ising model on a square lattice; the $J_{nm}$'s are the Ising coupling constants. Up arrows represent a metal in the HS state, and down arrows in the LS state.}
	\label{fig:ElasticToIsing}
\end{center}
\end{figure}

We consider a square lattice in two spatial dimensions, with the arrangement of metals and ligands depicted in Fig. \ref{fig:ElasticToIsing}; only nearest-neighbor elastic interactions, and a harmonic potential on the angle $\theta$ between adjacent metal-ligand bonds are included. Even though the following discussion focuses on this specific model on a square lattice, the method used can be readily extended to other elastic models and lattices. 

This model is appropriate for modeling the in-plane physics of frameworks \cite{Real,JaceSquare}, with the chemical formula $[M(L')_n(L)_2]$, where $M$ is the SCO active metal, $L$ is the in-plane ligand, $L'$ is the interlayer ligand, and $n=1$ or 2 for bridging and monodentate ligands respectively. The $L'$ ligand is not explicitly described by this model. A typical member of this family is [Fe(azpy)$_2$(NCS)$_2$], where azpy is \textit{trans}-4,4'-azopyridine \cite{Halder}. The model can also serve to describe  Hoffmann-like frameworks \cite{Ni,JaceSquare} with the generic formula $[M(L')_n\{M'(L)_2\}_2]$, where $M'$ is a non-SCO-active metal. A typical example is [Fe(pz)\{Au(CN)$_2$\}$_2$], where pz is pyrazine \cite{Ni}. In this case the `ligands' in the model actually describe the composite $M'(L)_2$ groups, e.g., Au(CN)$_2$ in [Fe(pz)\{Au(CN)$_2$\}$_2$].

The Hamiltonian describing this system is
\begin{eqnarray}\label{eq:HSquareElastics}
H &=& H_\sigma + K + V,
\end{eqnarray}
where the three terms are respectively, the free energy differences between the HS and LS states of  isolated SCO centers,
the kinetic energy of the metal ions and ligands, and the elastic energy of the system. Specifically:
\begin{eqnarray}
H_\sigma
&=&\frac{\Delta G}2\sum_{i\in\mathbb{Z}_{N_x}}\sum_{j\in\mathbb{Z}_{N_y}} 
\sigma_{i,j},
\\ 
K &=& 
\sum_{i\in\mathbb{Z}_{N_x}}\sum_{j\in\mathbb{Z}_{N_y}} \frac{\bm{p}_{M,i,j}^2}{2m_M} 
+ \frac{\bm{p}_{{L_1},i+\frac12,j}^2}{2m_{L_1}} + \frac{\bm{p}_{L_2,i,j+\frac12}^2}{2m_{L_2}},   
\\ 
V &=& \sum_{i\in\mathbb{Z}_{N_x}}\sum_{j\in\mathbb{Z}_{N_y}} \bigg[
 \frac{k}{2} \left( |\bm{r}_{{L_1},i+\frac12,j} - \bm{r}_{M,i,j}| - \overline{R} - \delta\sigma_{i,j} \right)^2 
\notag 
\\ && \hspace{1cm} 
+ \frac{k}{2} \left( |\bm{r}_{M,i,j} - \bm{r}_{{L_1},i-\frac12,j}| - \overline{R} - \delta\sigma_{i,j} \right)^2 
\notag 
\\ && \hspace{1cm} 
+ \frac{k}{2} \left( |\bm{r}_{{L_2},i,j+\frac12} - \bm{r}_{M,i,j}| - \overline{R} - \delta\sigma_{i,j} \right)^2 
\notag 
\\ && \hspace{1cm} 
+ \frac{k}{2} \left( |\bm{r}_{M,i,j} - \bm{r}_{{L_2},i,j-\frac12}| - \overline{R} - \delta\sigma_{i,j} \right)^2 
\notag 
\\ && \hspace{1cm} 
+\frac{k_\theta}{2} \overline{R}^2\sum_{n=1}^4   \sin^2\left( \frac{\pi}{2}-\theta_{i,j,n} \right) 
\bigg].
\label{eq:VSquareElastics}
\end{eqnarray}
We have assigned a pseudo-spin variable $\sigma$ to each metal atom, such that $\sigma=1$ if the metal is in the high spin (HS) state, and $\sigma=-1$ if it is in the low spin (LS) state. The volume of a SCO molecule decreases when going from the HS state to the LS state, as such  the minimum energy metal-ligand separation is dependent on the spin-state of the metal: we set $\overline{R}=(R_{HS}+R_{LS})/2$ and $\delta=(R_{HS}-R_{LS})/2$, where $R_{HS}$ ($R_{LS}$) is the equilibrium distance between the centre of a ligand and the center of a nearest neighbor metal atom in the HS (LS) state. $\Delta G=\Delta H-T\Delta S$ is the free energy difference between the HS and LS states of a single metal center, where $\Delta H$ and $\Delta S$ are the enthalpy and entropy differences between the HS and LS states of a single metal center, and $T$ is temperature, $\bm{r}_{\nu,i,j}$ and $\bm{p}_{\nu,i,j}$ are the position and momentum of a group, where $\nu=M$, ${L_1}$, or ${L_2}$ labels the species of the group. Note that $i$ and $j$ are both integers for the metals, but one of them is a half-odd-integer for every ligand. $m_\nu$ is the mass of species $\nu$, $k$ is the elastic constant of the springs,  and $N=N_x\times N_y$, is the number of unit cells. 

We define
\begin{equation}
	\bm{r}_{\nu,i,j} = \bm{a}_{i,j} +\bm{u}_{\nu,i,j} . 
\end{equation}
where $\bm{u}_{\nu,i,j}=(u^{(x)}_{\nu,i,j},u^{(y)}_{\nu,i,j})$ describes displacements about
\begin{equation}
    \bm{a}_{i,j}=2ia_x\bm{\hat{x}}+2ja_y\bm{\hat{y}}, 
\end{equation}
$a_x$ ($a_y$) are the average metal-ligand distances in the $x$ ($y$) directions, and $\bm{\hat{x}}$ ($\bm{\hat{y}}$) are unit vectors in the $x$ ($y$) directions. Alternatively, we can observe that the lattice constants are $2a_x$ and $2a_y$.

\begin{figure*}
	\begin{center}
		\includegraphics[width=1.9\columnwidth]{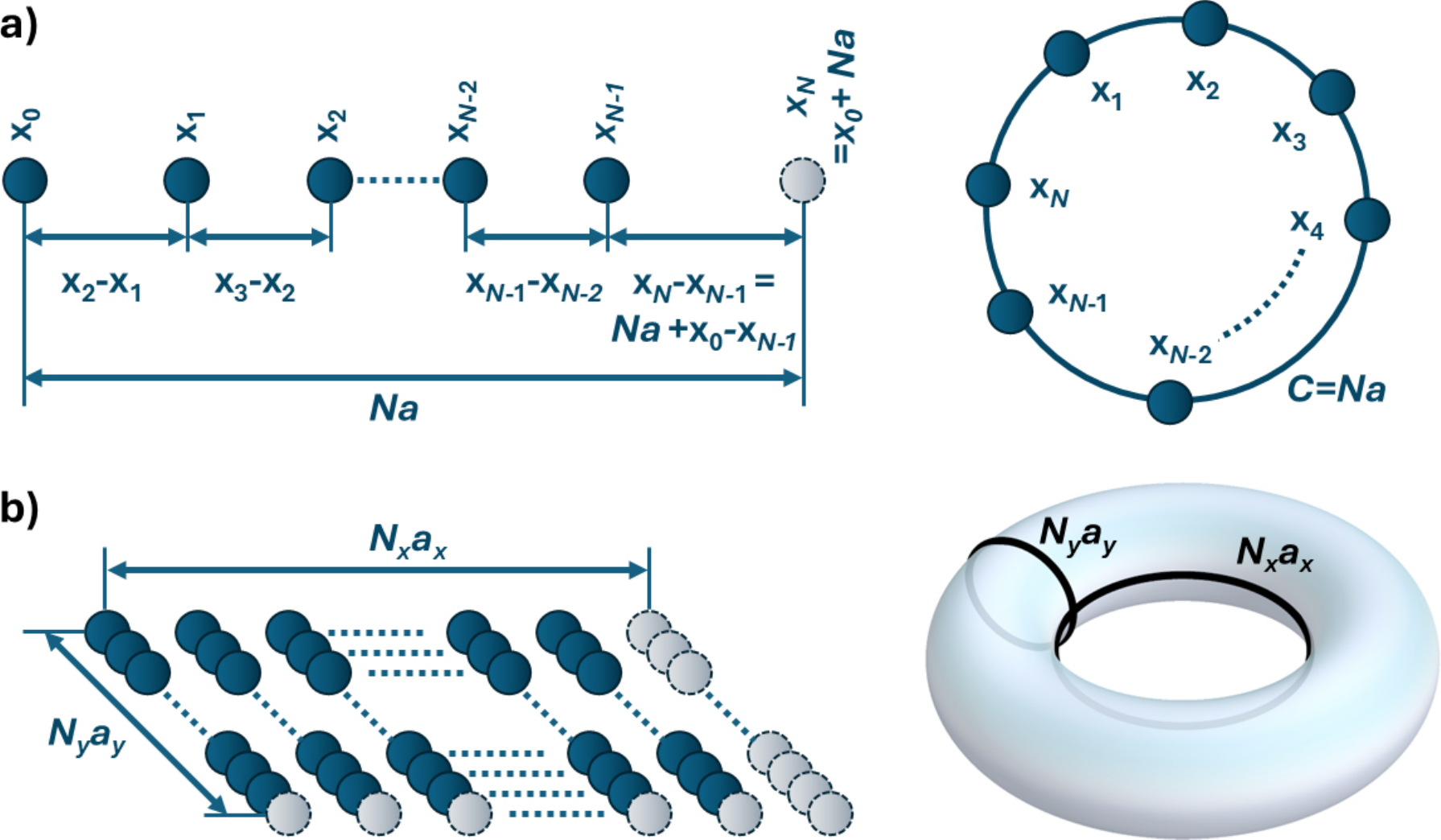} 
		\caption{Counting degrees of freedom with periodic boundary conditions (PBCs). To simplify these sketches we consider only one type of particle per unit cell, with an average spacing $a$ (or $a_x$ and $a_y$ in two-dimensions). (a) In one-dimension particles move on a line. PBCs produce a copy of the zeroth particle in position $N$, where $N$ is the  number of particles. The location of particle $N$ is given by $x_{N}=x_0+Na$. Therefore $a$ is independent of the $N$ particle coordinates for systems with PBCs and must be included explicitly as a variable. Thus there are $N+1$ structural parameters. This problem is topologically equivalent to particles moving on the perimeter of a circle. In this picture $a$ determines the circumference of the circle, $C=Na$), thus we  have $N+1$ structural degrees of freedom, as expected. (b) In two dimensions for particles moving in a plane there are $N+2$ structural degrees of freedom: the $x$ and $y$ coordinates of the $N=N_xN_y$ particles plus $a_x$ and $a_y$. Note that  the unit cell angle is not an independent parameter. This is clear if we map the plane with PBCs to a torus, where there are two independent circumferences, $N_xa_x$ and $N_ya_y$, in additions to the $N$ coordinates of the particles. In $d$ dimensions there are $N+d$ independent coordinates.}
		\label{fig:PBC}
	\end{center}
\end{figure*}

It will be useful below to define 
\begin{equation}
a_\alpha=\overline{R}+\epsilon_\alpha, 
\end{equation}
where  $\alpha\in\{x,y\}$.
This   simplifies the  treatment of the expansion and contraction of the lattice, which we will see below plays a crucial role in SCO. 
For periodic boundary conditions, which we use throughout this paper, this construction has the required number of structural variables (as the $a_\alpha$ determine the size of the periodic cell for fixed $N_x$, $N_y$; see Fig. \ref{fig:PBC}).

\subsection{Normal modes}

The harmonic approximation, $|\bm{u}_{M,i,j} - \bm{u}_{{L_1},i-\frac12,j}| \ll a_x$, is implicit in any harmonic elastic model. Therefore, without introducing any approximations beyond those already in the model we can write
\begin{widetext}
\begin{eqnarray}
|\bm{r}_{M,i,j} - \bm{r}_{{L_1},i-\frac12,j}| - \overline{R} - \delta\sigma_{i,j}
	&=& u^{(x)}_{M,i,j} - u^{(x)}_{{L_1},i-\frac12,j} + \epsilon_x - \delta\sigma_{i,j}.  \label{eq:harmonic}
\end{eqnarray}
Similar results follow for $|\bm{u}_{M,i,j} - \bm{u}_{{L_2},i,j-\frac12}|$, $|\bm{u}_{{L_1},i+\frac12,j}-\bm{u}_{M,i,j}|$, and $|\bm{u}_{{L_2},i,j+\frac12}-\bm{u}_{M,i,j}|$.
Hence, 
\begin{eqnarray}
V &=& V_u+V_{u\sigma} + V_{\epsilon \sigma},
\label{eq:V in u epsilon}
\end{eqnarray}
where
\begin{eqnarray}
V_u &=& \sum_{i,j} \left[  
\frac{k}{2} \left(  u^{(x)}_{{L_1},i+\frac12,j} -  u^{(x)}_{M,i,j} \right)^2
+ \frac{k}{2} \left(  u^{(x)}_{M,i,j} -  u^{(x)}_{{L_1},i-\frac12,j} \right)^2 
\notag \right.
\\ && \hspace{0.6cm} \left.
+ \frac{k}{2} \left(  u^{(y)}_{{L_2},i,j+\frac12} -  u^{(y)}_{M,i,j} \right)^2 
+ \frac{k}{2} \left(  u^{(y)}_{M,i,j} -  u^{(y)}_{{L_2},i,j-\frac12} \right)^2 
\notag \right.
\\ && \hspace{0.6cm} \left.
+\frac{k_\theta}{2}\left(u^{(x)}_{{L_2},i,j+\frac12} - u^{(x)}_{M,i,j} + u^{(y)}_{{L_1},i+\frac12,j} - u^{(y)}_{M,i,j} \right)^2
+ \frac{k_\theta}{2}\left(u^{(x)}_{M,i,j} - u^{(x)}_{{L_2},i,j+\frac12} + u^{(y)}_{{L_1},i-\frac12,j} - u^{(y)}_{M,i,j} \right)^2
\notag \right.
\\ && \hspace{0.6cm} \left.
+ \frac{k_\theta}{2}\left(u^{(x)}_{{L_2},i,j-\frac12} - u^{(x)}_{M,i,j} + u^{(y)}_{{L_1},i-\frac12,j} - u^{(y)}_{M,i,j} \right)^2
+\frac{k_\theta}{2}\left(u^{(x)}_{{L_2},i,j-\frac12} - u^{(x)}_{M,i,j} + u^{(y)}_{M,i,j} - u^{(y)}_{{L_1},i+\frac12,j} \right)^2\right],
\label{eq:Vu1}
\end{eqnarray}
\begin{eqnarray}
V_{u\sigma} &=& k\delta\sum_{i,j} \left[ u^{(x)}_{{L_1},i-\frac12,j}-u^{(x)}_{{L_1},i+\frac12,j}
+u^{(y)}_{{L_2},i,j-\frac12}-u^{(y)}_{{L_2},i,j+\frac12}\right]\sigma_{i,j}
\end{eqnarray}
and
\begin{eqnarray}
V_{\epsilon\sigma} &=& 
k \sum_{i,j} \left[\left( \epsilon_x - \delta\sigma_{i,j} \right)^2 + \left( \epsilon_y - \delta\sigma_{i,j} \right)^2\right]
=
N k (\epsilon_x^2  + \epsilon_y^2)
- 2 k\delta (\epsilon_x + \epsilon_y) \sum_{i,j} \sigma_{i,j} 
+2 N k \delta^2
\end{eqnarray}
as $\sigma_{i,j}^2=1$.

Next, we Fourier transform the generalized coordinates, 
\begin{eqnarray}
u^{(\alpha)}_{M,i,j}&=&\frac{1}{\sqrt{N}}\sum_{\bm{q}}e^{i\bm{q}\cdot\bm{n}_{i,j}}u^{(\alpha)}_{M,\bm{q}},\\
u^{(\alpha)}_{{L_1},i+\frac12,j}&=&\frac{1}{\sqrt{N}}\sum_{\bm{q}}e^{i\bm{q}\cdot\bm{n}_{i+\frac12,j}}u^{(\alpha)}_{L_1,\bm{q}},\\
u^{(\alpha)}_{{L_2},i,j+\frac12}&=&\frac{1}{\sqrt{N}}\sum_{\bm{q}}e^{i\bm{q}\cdot\bm{n}_{i,j+\frac12}}u^{(\alpha)}_{L_2,\bm{q}},
\end{eqnarray}
and the pseudospin variables
\begin{equation}
\sigma_{i,j} = \frac{1}{\sqrt{N}} \sum_{\bm{q}} e^{i\bm{q}\cdot\bm{n}_{i,j}} \sigma_{\bm{q}},
\label{Eq: FT sigma}
\end{equation}
where we have introduced the shorthand notation $\bm{n}_{i,j}=(i,j)$ and ${\bm{q}}=2\pi(m_x/N_x,m_y/N_y)$ with $m_\alpha\in\mathbb{Z}_{N_\alpha}$.
This yields
\begin{eqnarray}
K &=& \sum_{\bm{q}} \bigg[  \frac{\bm{p}_{M,\bm{q}}\cdot\bm{p}_{M,-\bm{q}}}{2m_M} 
+\frac{\bm{p}_{L_1,\bm{q}}\cdot\bm{p}_{L_1,-\bm{q}}}{2m_{L_1}} 
+\frac{\bm{p}_{L_2,\bm{q}}\cdot\bm{p}_{L_2,-\bm{q}}}{2m_{L_2}} 
 \bigg],
 \\
	V_{\epsilon\sigma} &=& 
	N k \left( {\epsilon}_x^2 + {\epsilon}_y^2 \right)
	- 2 {N}  k\delta (\epsilon_x +\epsilon_y) \langle\sigma\rangle
	+2 N k \delta^2,
	\label{eq:Vepssing-int}
\end{eqnarray}
where 
\begin{equation}
	\langle\sigma\rangle = \frac1{N} \sum_{i,j} \sigma_{i,j} = \frac{\sigma_{\bm0}}{\sqrt{N}}
	\label{eq:sigma0}
\end{equation}
is the average of the $\sigma_{i,j}$,
and
\begin{eqnarray}
V_u &=& \sum_{\bm{q}} \bigg[
ku^{(x)}_{L_1,\bm{q}}u^{(x)}_{L_1,-\bm{q}}
-2k\cos(q_x)u^{(x)}_{M,\bm{q}}u^{(x)}_{L_1,-\bm{q}}
+(k+2k_\theta)u^{(x)}_{M,\bm{q}}u^{(x)}_{M,-\bm{q}} 
+ ku^{(y)}_{L_2,\bm{q}}u^{(y)}_{L_2,-\bm{q}}
\notag
\\ && \hspace{0.7cm}
-2k\cos(q_y)u^{(y)}_{M,\bm{q}}u^{(y)}_{L_2,-\bm{q}}
+2k_\theta u^{(x)}_{L_2,\bm{q}}u^{(x)}_{L_2,-\bm{q}}
+(k+2k_\theta)u^{(y)}_{M,\bm{q}}u^{(y)}_{M,-\bm{q}}
+2k_\theta u^{(y)}_{L_1,\bm{q}}u^{(y)}_{L_1,-\bm{q}}
\notag
\\ && \hspace{0.7cm}
-4k_\theta\big(\cos(q_y)u^{(x)}_{M,\bm{q}}u^{(x)}_{L_2,-\bm{q}}
+\cos(q_x)u^{(y)}_{M,\bm{q}}u^{(y)}_{L_1,-\bm{q}}
-\sin(q_x)\sin(q_y)u^{(x)}_{L_2,\bm{q}}u^{(y)}_{L_1,-\bm{q}}\big)\bigg].
\end{eqnarray}
\end{widetext}
We rewrite $V_u$ as
\begin{equation}\label{eq:Hu}
    V_{u}=\frac{1}{2}\sum_{\bm q,\alpha,\alpha^\prime,\nu,\nu^\prime}\left(\mathrm{f}^{\alpha,\alpha^\prime}_{\nu,\nu^\prime,\bm{q}}u^{(\alpha)}_{\nu,\bm{q}}u^{(\alpha^\prime)}_{\nu^{\prime},-\bm{q}}\right)\equiv\frac12\sum_{\bm q}\bm u_{\bm{q}}^T\mathrm{F}_{\bm{q}}\bm u_{\bm{-q}},
\end{equation}
where $\mathrm{F}_{\bm{q}}$ is the Hessian matrix whose elements are 
\begin{equation}\label{eq:hessianelements}
    \mathrm{f}^{\alpha,\alpha^\prime}_{\nu,\nu^\prime,\bm{q}}=\frac{\partial^2V_u}{\partial u^{(\alpha)}_{\nu,\bm{q}}\partial u^{(\alpha^\prime)}_{\nu^\prime,\bm{-q}}}.
\end{equation}

We diagonalize $V_u$ by writing the eigenvalue problem
\begin{equation}
    \kappa_{\bm{q}}\bm{\psi}_{\bm{q}}=\mathrm{F}_{\bm{q}}\bm{\psi}_{\bm{q}},
\end{equation}
and find the eigenvalues, $\kappa_{\bm{q}}$, and normalized eigenvectors, $\bm{\psi}_{\bm{q}}$, such that
\begin{equation}
    \bm u_{\bm{q}}^T\mathrm{F}_{\bm{q}}\bm u_{\bm{-q}}=\bm{Q}^T_{\bm{q}}D_{\bm{q}}\bm{Q}_{\bm{-q}} ,  
\end{equation}
where $\bm{Q}_{\bm{q}}=U^T_{\bm{q}}\bm{u}_{\bm{q}}$, $\bm u_{\bm{q}}$ is a six-dimensional column vector whose components are $u^{(\alpha)}_{\nu,\bm{q}}$, $U_{\bm{q}}$ is a square matrix whose columns are $\bm{\psi}_{\bm{q}}$, and $D_{\bm{q}}$ is a diagonal matrix whose elements are $\kappa_{\bm{q}}$. With this transformation $u^{(\alpha)}_{\nu,i,j}$ takes the form
\begin{eqnarray}\label{eq:general-solution}
u^{(\alpha)}_{\nu,i,j} = \frac{1}{\sqrt{N}} \sum_{\mu,\bm{q}} e^{i\bm{q}\cdot\bm{n}_{i,j}} Q_{\mu,\bm{q}}\psi^{(\alpha)\mu}_{\nu,\bm{q}},
\label{eq:u-from-Q}
\end{eqnarray}
where $\mu \in [1,6]$ enumerates the eigenvalues of the Hessian matrix, $Q_{\mu,\bm{q}}$ are the components of the column vector $\bm{Q}_{\bm{q}}$, and $\psi^{(\alpha)\mu}_{\nu,\bm{q}}$ are the components of the orthonormal eigenvectors $\bm{\psi}^{\mu}_{\bm{q}}$. With this change of basis 
\begin{equation}
    V_u = \frac{1}{2}\sum_{\mu,\bm{q}}\kappa_{\mu,\bm{q}}Q_{\mu,\bm{q}}Q_{\mu,\bm{-q}}\label{eq:vibrational-H}
\end{equation}
and
\begin{equation}
    V_{u\sigma} = \sum_{\mu,\bm{q}} \xi_{\mu}(\bm{q}) Q_{\mu,\bm{q}} \sigma_{-\bm{q}},
\end{equation}
where
\begin{eqnarray}
\xi_{\mu}(\bm{q}) &=& -2i\delta k\left[ \psi^{(x)\mu}_{L_1,\bm{q}}\sin(q_x)  +\psi^{(y)\mu}_{L_2,\bm{q}}\sin(q_y)\right].
\end{eqnarray}

\subsection{Displaced oscillator transformations}

We make a displaced oscillator transformation, which defines
\begin{eqnarray}
v_{\mu,\bm{q}} &=& Q_{\mu,\bm{q}} + \lim_{\eta\rightarrow 0} \left(\frac{\xi_{\mu}(-\bm{q})}{\kappa_{\mu,\bm{q}} + \eta} \right)\sigma_{\bm{q}}.
\label{eq:disp-v}
\end{eqnarray}
This is essentially ``completing the square'' so that the interaction between $Q$ and $\sigma$ is replaced by an interaction between pseudo-spins. This is mathematically analogous to treatments of the co-operative Jahn-Teller effect \cite{Kanamori,Gehring_1975}.
Similarly, we  make displaced oscillator transformations for the changes in the lattice constants:
	\begin{eqnarray}
		\varepsilon_\alpha &=& 
		\epsilon_\alpha -  \delta \langle\sigma\rangle, \label{eq:disp-eps}
	\end{eqnarray}
which yield
\begin{eqnarray}
V_{\epsilon\sigma} 
&=& 
Nk \left[ \left( \varepsilon_x^2 + \varepsilon_y^2 \right) 
- 2 \delta^2 \langle\sigma\rangle^2
+2  \delta^2\right].
\label{eq:Vepssig}
\end{eqnarray}
Note that the second term
\begin{eqnarray}
		- 2 N k \delta^2 \langle\sigma\rangle^2 &=& 
	- 2N k \delta^2 (2\gamma_{HS}-1)^2 \notag \\
	&=& - \frac{2 k \delta^2}N \sum_{i,i',j,j'}\sigma_{i,j}\sigma_{i',j'}.
\end{eqnarray}
where $\gamma_{HS}=\sum_{i,j}(1+\sigma_{i,j})/2N
=[1+\langle\sigma\rangle]/2$ is the fraction of HS metal centers.
Hence the displaced oscillator transformation has replaced the interaction between the average spin state and the lattice constants with an Husimi-Temperley interaction between spin states that is independent of the separation of the metal centers, as it is independent of the separation of the metal centers, yet extensive.

\begin{figure}
	\begin{center}
		\includegraphics[width=0.8\columnwidth]{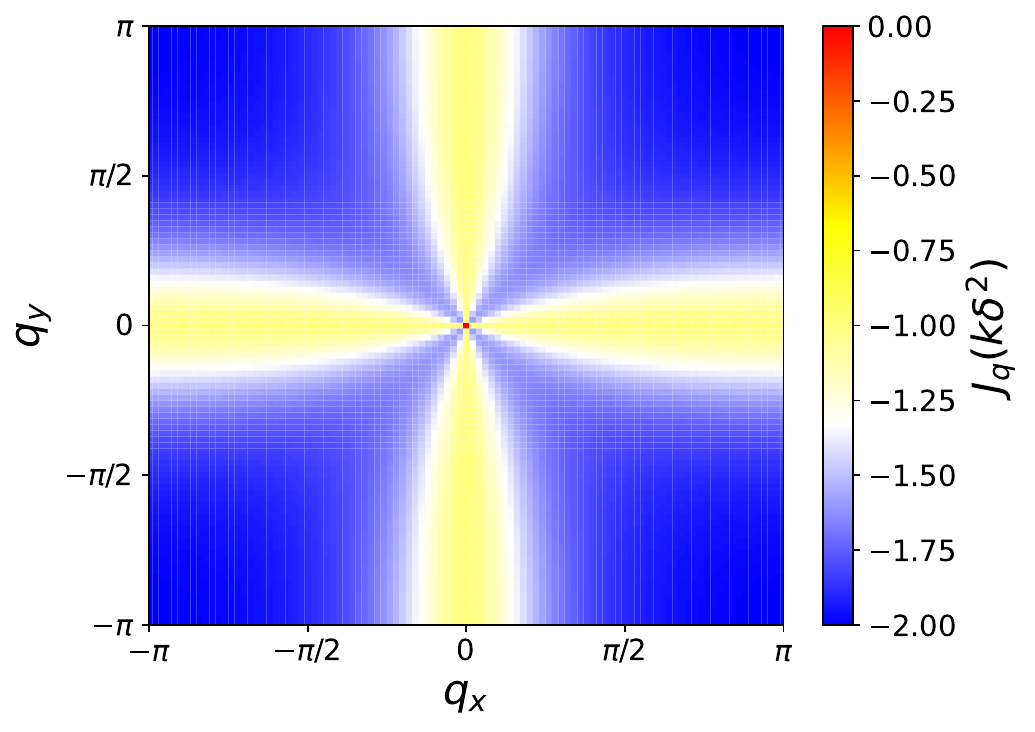} 
		\caption{Wavevector dependence of the Ising interactions, $J_{\bm q}/k\delta^2$ for $k_\theta / k =1/15$. The interactions are strongest along the diagonals ($q_x=\pm q_y$) and vanish only exactly at the Brillouin zone center (${\bm q}={\bm 0}$). }
		\label{fig:Jq}
	\end{center}
\end{figure}

Finally, combining all the terms, the potential (Eq. \ref{eq:V in u epsilon}) can be written as
\begin{eqnarray}
 	V &=& V_{v, \epsilon} + H_\sigma + H_{\sigma\sigma},
	\label{Eq: Hq final}
\end{eqnarray}
where
\begin{eqnarray}
	V_{v,\epsilon}  &=&  \frac12\sum_{\mu,\bm{q}}\kappa_{\mu,\bm{q}} v_{\mu,\bm{q}} v_{\mu,-\bm{q}}  
	+ N k \left(\varepsilon_x^2 +  \varepsilon_y^2 \right)
	, 
	\label{eq:Hnu-final}
\end{eqnarray}
and
\begin{eqnarray}
	H_{\sigma\sigma} &=&  2N k \delta^2\left(1-\langle\sigma\rangle^2\right) + \sum_{\bm{q}} J_{\bm q} \sigma_{\bm{q}}\sigma_{-\bm{q}}
	\notag\\
	&=& 8 N k \delta^2 \gamma_{HS}\left(\gamma_{HS}-1\right) + \sum_{\bm{q}} J_{\bm q} \sigma_{\bm{q}}\sigma_{-\bm{q}},
	\label{eq: Hsigmasigma}
\end{eqnarray}
with
\begin{eqnarray}
	J_{\bm{q}} &=& - \sum_{\mu} \kappa_{\mu,\bm{q}} 
	\lim_{\eta\rightarrow 0} \left( \frac{|\xi_{\mu}(\bm{q})|^2}{2(\kappa_{\mu,\bm{q}} + \eta)^2} \right)
	\label{eq:Jq},
\end{eqnarray}
which we plot  in Fig. \ref{fig:Jq}. A detailed discussion of $J_{\bm{q}}$ is deferred until section \ref{sect:long_range}.

$V_{v,\epsilon}$ is independent of the $\sigma_{i,j}$ (Eq. (\ref{eq:Hnu-final})). Whereas $H_\sigma$ and $H_{\sigma\sigma}$ are independent of the the $v_{\mu,\bm{q}}$ and $\varepsilon_\alpha$. Therefore the  displaced oscillators  are decoupled from the pseudospins.

\subsection{Exact ground states} \label{sec: Exact ground states}

It immediately follows from the definition of the Fourier transform of the pseudospins (Eq. \ref{Eq: FT sigma}) that
since $i$ and $j$ are integers for the metals

\begin{eqnarray}
	\sigma_{ \bm \pi } 
	&=& 
	\frac{1}{\sqrt{N}}  \sum_{i,j} (-1)^{i+j} \sigma_{i,j},
\end{eqnarray}
where we have introduced the shorthand ${\bm\pi}=\left(\pi, \pi \right)$.
Thus $\sigma_{ \bm{0}} $ (see Eq. (\ref{eq:sigma0})) and $\sigma_{ \bm\pi }$ are proportional to the pseudospin-magnetization and staggered-pseudospin-magnetization respectively.
For the ideal  ferroelastic states ($\sigma_{i,j}=1$ for all $i, j$ or $\sigma_{i,j}=-1$ for all $i, j$) one has  $\sigma_{\bm{0}} = \pm\sqrt{N} $ and $\sigma_{\bm{q}} = 0$ for all $\bm{q}\ne\bm0$.
For the ideal checkerboard states (with all nearest neighbors of each HS ion in the LS state and \textit{vice versa}) one finds $\sigma_{ \bm\pi } = \pm\sqrt{N}$ and $\sigma_{\bm{q}} = 0$ for all $\bm{q}\ne\bm\pi$.
Since $J_{\bm{0}}=0$ and $J_{\bm\pi}=  - 2 k\delta^2 $ (cf. Fig. \ref{fig:Jq}), the Ising interaction term of the Hamiltonian (Eq.  \ref{eq: Hsigmasigma}) gives the same energy for both the ideal ferroelastic and the ideal checkerboard states: $H_{\sigma\sigma}=0$. 

The  potential energies  (Eq. \ref{Eq: Hq final}) for the ideal ferroelastic  and  chequerboard states are thus
\begin{eqnarray}
	V &=&  V_{v, \epsilon} + H_\sigma,
\end{eqnarray}
Clearly, $\min (V_{v, \epsilon})=0$, which occurs for $v_{\mu,\bm{q}} = \varepsilon_x =  \varepsilon_y=0$ for all $\bm{q}$ (cf. Eq. \ref{eq:Hnu-final}). Therefore, the ideal ferroelastic and checkerboard states are degenerate for $\Delta G=0$ and both have $V=0$. It is straightforward to confirm this result from the elastic Hamiltonian Eqs. (\ref{eq:HSquareElastics}-\ref{eq:VSquareElastics}) as $K$ and $V$ are positive semi-definite. This also implies that these two states are ground states of the model. 

The above result does not preclude other possible degenerate ground states. However, as we find numerically that $J_{\bm q}>-2k\delta$ for all ${\bm q}\ne{\bm \pi}$ (see Fig. \ref{fig:Jq}), we  conclude that there are no other  ground states with long-range order. 
For $\Delta G\ne0$ either the all-HS ($\Delta G<0$) or the all-LS ($\Delta G>0$) states becomes a unique ground state.

%


\subsection{
	Ising-Husimi-Temperley model}\label{sect:long_range}
	
Fourier transforming the spin Hamiltonian Eq. (\ref{eq: Hsigmasigma})  to real space, using Eq. (\ref{Eq: FT sigma}), gives
\begin{equation}
	H_{\sigma\sigma} = \sum_{i,j} \sum_{n,m} J_{n,m} \sigma_{i,j}\sigma_{i+n,j+m} 
	+ \frac{ J_{\infty,\epsilon} }{N} \sum_{i,j,i',j'}  \sigma_{i,j} \sigma_{i',j'} ,
\end{equation}
where
\begin{eqnarray}
	J_{n,m} &=&  \frac{1}{N}  \sum_{\bm{q}}     J_{\bm q}  e^{i\bm{q}\cdot \bm{n}_{n,m} }
	\label{eq: Jnm}
\end{eqnarray}
and
\begin{eqnarray}
	J_{\infty,\epsilon} &=&  -2 k\delta^2 .
	\label{eq: Jinfep}
\end{eqnarray}

\begin{figure}
	\begin{center}
		\includegraphics[width=0.8\columnwidth]{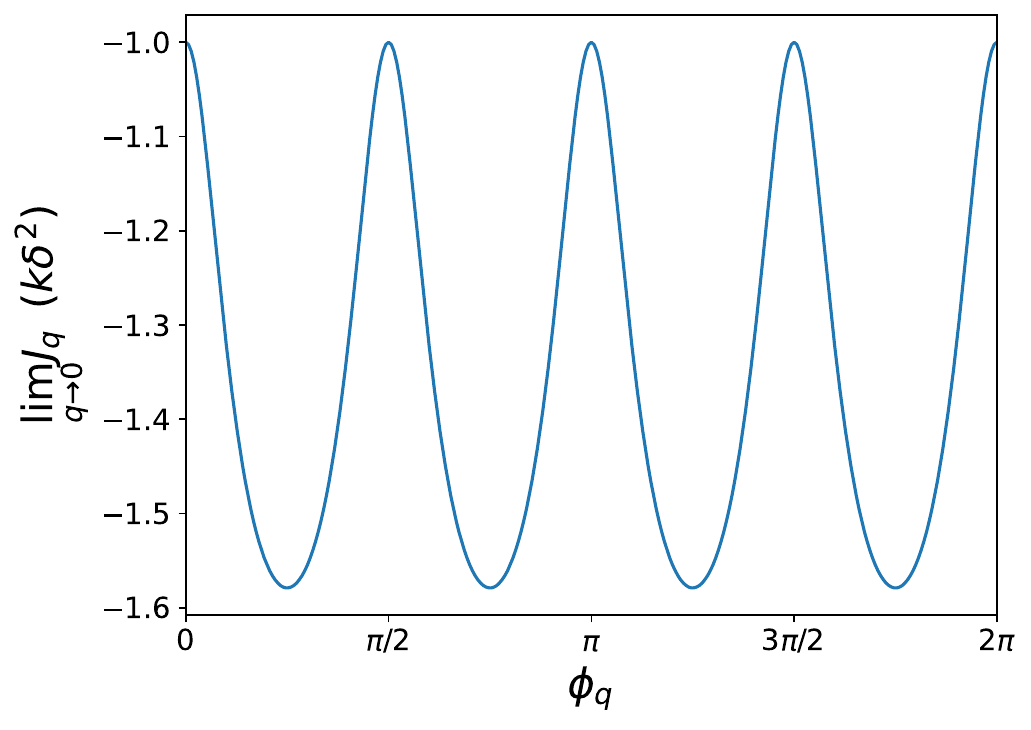} 
		\caption{Numerical evidence that the small wavevector Ising interaction strength, $\lim_{q\to0}J_q $ does not vanish and depends on the angle of approach to the Brillouin zone center, $\phi_{q}$. Here we plot $\lim_{{\bm q}\to{\bm 0}}J_{\bm q} $ for $k_\theta / k =1/15$, at $|\bm q| = \pi\times10^{-5} $. }
		\label{fig: JqPhi}
	\end{center}
\end{figure}

In the thermodynamic limit $\bm q$ goes from a discrete set to a continuum and the sum over $\bm q$ in Eq. (\ref{eq: Jnm}) changes to an integral over the Brillouin zone.
Mathematically, taking the continuum limit requires that the function $J(\bm q)$ must satisfy $J(\bm q) = \lim_{\bm{q'}\to \bm q} J(\bm q')$ for all $\bm q$ in the discrete domain. However, at the Brillouin zone center ($\bm{q}=\bm{0}$) $J_{\bm{q}}$  is not smooth: it is a punctured function \cite{stein2010complex}  as  $J_{\bm{0}}=0$ and $\lim_{\bm{q}\to \bm{0} }J_{\bm{q}} \neq 0$, as shown in Fig. \ref{fig:Jq}. 
Furthermore, in general, the limit depends on the angle of approach  [$\phi_q$, where $(q_{x},q_{y})=q(\cos\phi_q, \sin\phi_q)$] to the Brillouin zone center, as shown in Fig. \ref{fig: JqPhi}.

Therefore, before taking the continuum limit, we must transform the effective Ising interaction to a smooth function. Due to the angular dependence of the limit, fixing the puncture is simpler in polar coordinates, thus we transform $J_{\bf{q}}$ to polar coordinates ${\bm q}=(q, \phi_q)$ and add a delta function to define
\begin{eqnarray}
	J^\text{SR}_{(q,\phi_q)} &\equiv&  J_{(q,\phi_q)}  -  \delta_{{q},{0}} \left(J_{\bm 0} - \lim_{{q}\to{0}} J_{(q,\phi_q)}\right).
\end{eqnarray}
This ensures that $J^\text{SR}_{(q,\phi_q)}$ is a smooth function. 
In real space [Eq. (\ref{eq: Jnm})]
\begin{eqnarray}
	J_{n,m} &=&  \frac{1}{N}  \sum_{\bf{q}}  J^\text{SR}_{\bf{q}}  e^{-i\bm{q}\cdot \bm{n}_{n,m} } \notag \\
    && + \frac{1}{N}   \sum_{\phi_q}  \left(J_{\bm 0} - \lim_{{q}\to{0}} J_{(q,\phi_q)}\right). 
\end{eqnarray}
Note that for the model studied here $J_{\bm 0}=0$, but we retain the constant for the sake of generality.

It is now straightforward to take thermodynamic limit  and convert discrete sums over $\bf{q}$ into the integrals over the Brillouin zone.
\begin{eqnarray}
	J_{n,m} &=&  \frac{1}{4\pi^2}  \iint_{-\pi}^\pi d{\bf q}\  J^{\text{SR}}_{\bf{q}}  e^{-i\bm{q}\cdot \bm{n}_{n,m}} \notag \\
    && + \frac{1}{N} \left(J_{\bm 0} -  \frac{1}{2\pi} \int_{\phi_q}  d\phi_q \lim_{{q}\to{0}} J_{(q,\phi_q)}\right).
\end{eqnarray}
Finally we have 
\begin{eqnarray}\label{eq:finalHamiltonian}
    H&=&\frac{1}{2}\sum_{i,j}\left(\Delta H-T\Delta S\right)\sigma_{i,j}+\sum_{i,j,n,m}J_{n,m}^\text{SR}\sigma_{i,j}\sigma_{i+n,j+m}
    \notag \\ 
  	&&+  \frac{ J_{\infty}}{N}   \sum_{i,j,i',j'}  \sigma_{i,j} \sigma_{i',j'}
   	+  H_{v},
  	 \label{eq:Hgamma}
\end{eqnarray}
where 
\begin{eqnarray}
	J_{n,m}^\text{SR} &=&  \frac{1}{4\pi^2} \iint_{-\pi}^\pi d{\bf q}\  J^{\text{SR}}_{\bf{q}}  e^{-i\bm{q}\cdot \bm{n}_{n,m}}
	\label{eq:IntegralJnmSR}
\end{eqnarray}
are the Ising interactions between pseudospins, 
\begin{equation}
	J_{\infty} = J_{\infty,\epsilon} +  J_{\infty,u}   
\end{equation} 
is a Husimi-Temperley interaction that couples all pseudospins to each other with the same strength regardless of their physical separation.
$J_{\infty,\epsilon}$ arises from the coupling of the pseudospins to the changes in the lattice constants ($\epsilon_x$ and $\epsilon_y$) while 
\begin{equation}
J_{\infty,u} \equiv J_{\bm 0} -  \frac{1}{2\pi} \int_{\phi_q}  d\phi_q \lim_{{q}\to{0}} J_{(q,\phi_q)}.
\end{equation}
arises from the coupling of the pseudospins to changes in the positions of the ligands.

The presence of Husimi-Temperley interaction due to fluctuations in the lattice constants was anticipated in related models in the limit $k_\theta\rightarrow\infty$ \cite{JaceKagome,JaceSquare,JacePyro,Isomophism}. 
In this limit, in models with spring constants, $k_{n,m}$, directly between metal centers  $J_{\infty,\epsilon}=-\sum_{n,m} J_{n,m}=-\delta^2\sum_{n,m}k_{n,m}\eta_{n,m}$ \footnote{Note that $J_{\infty}$ is defined with the opposite sign in Refs. \cite{JaceKagome,JaceSquare,JacePyro}.}, where $\eta_{n,m}$ is the dimensionless separation of the two metal site: $\eta_{n,m}=\sqrt{n^2+m^2}$ on the square lattice. In our model there are four nearest neighbors and the effective spring constant (of the two springs connecting neighboring metals via a ligand) is $k_{\pm1,0}=k_{0,\pm1}=k/2$. Hence, in the $k_\theta\rightarrow\infty$ limit one has $J_{\infty,\epsilon}=-2k\delta^2$, consistent with our general result (Eq. \ref{eq: Jinfep}). 
	
The method and results presented here is more general and  demonstrates that $J_{\infty,\epsilon}$ is independent of $k_\theta$ for the model considered here. However, the $k_\theta\rightarrow\infty$ result has a much simpler derivation. This allows us to see that this Husimi-Temperley interaction results from the self-interaction of the average spin state  via the elastic coupling. This can also be seen in the present calculation. Briefly, the equilibrium lattice constants are determined by minimizing $V_{\epsilon,\sigma}$  with respect to $\epsilon_x, \epsilon_y$, which gives $\epsilon_x=\epsilon_y=\delta\langle\sigma\rangle$ (trivially from Eq. \ref{eq:Vepssing-int}). Substituting this back into the expression for $V_{\epsilon,\sigma}$ (Eq. \ref{eq:Vepssing-int}) or, equivalently making the displaced oscillator transformation (see Eq. \ref{eq:Vepssig}), yields the term proportional to $\langle\sigma\rangle^2=\sum_i\sigma_i\sum_j\sigma_j/N^2$, i.e., the Husimi-Temperley interaction.
	 
The emergence of Husimi-Temperley interactions in the real space Hamiltonian is consistent with previous results for  functions with punctures at $\bm{q}\to\bm{0}$ \cite{FrechettePRB,Frechette1,Dantchev}. 
Most relevantly, a similar discontinuity in $J(q)$ at $\bm q=\bm 0$ was pointed out by Frechette \textit{et al}. \cite{Frechette1,FrechettePRB} for a similar elastic model on an isotropic triangular lattice. Elasticity theory on a system with $C_{6v}$ symmetry is isotropic \cite{Landau}, therefore they do not observe the anisotropy that we do for the square lattice (which has $C_{4v}$ symmetry). 

However, we are not aware of previous work that treats both $J_{\infty,\epsilon}$ and  $J_{\infty,u}$ on an equal footing, as we do here.

The equivalence of the short-range elastic model and the Ising-Husimi-Temperley model provides a natural explanation of previous calculations reporting that elastic models of SCO are in the mean-field universality class \cite{Miyashita,Frechette1}.



\subsection{Some analytically tractable cases}\label{sect:anal}

The net pairwise interaction, $J_{n,m}^\text{SR}+(J_\infty/N)$, can be evaluated exactly in a few special cases.

For a one-dimensional chain in one spatial dimension (and therefore with no bending possible), the eigenvalues and eigenvectors of the Hessian take simple analytical forms:
\begin{eqnarray}
	\kappa_{\pm,q}&=&2k\left[1\pm\cos(qa)\right],\\
	\bm{\psi}_{\bm{q}}^\pm &=& \frac{1}{\sqrt{2}}\left(\begin{array}{c}
		1 \\ \pm1
	\end{array}\right).
\end{eqnarray}
This yields $J_{\infty,\epsilon}=-k\delta^2$, $J_{q}=-k\delta^2$ for all $q\ne0$ but $J_{q=0}=0$. Hence, $J_{\infty,u}=J_{q=0}-\lim_{q\rightarrow0}J_q=k\delta^2$. Therefore, $J_\infty=J^\text{SR}_{n}=0$ for all $n$. 
Thus, there are no Ising-Husimi-Temperley interactions between the metal centers.
Crucially, neglecting either $J_{\infty,\epsilon}$ or $J_{\infty,u}$ would introduce a spurious Husimi-Temperley interaction, underlining that both must be treated correctly.

The vanishing of the Ising-Husimi-Temperley interactions is simple to understand in the original elastic model: the chain will simply increase or decrease in length to accommodate the metal centers with different spin states. This result is  consistent with the one-dimensional model presented by Boukheddaden, \textit{et. al.}~\cite{Bouk1D}. In their model, spin crossover molecules are connected by springs whose elastic constants depend on the spin states of the molecules. In the case where all elastic constants are equal the Ising-Husimi-Temperley coupling constants vanish.

For $k_\theta=0$ the two-dimensional problem reduces to two uncoupled copies of the one-dimensional problem (this can be seen from Eq. (\ref{eq:Vu1})).
Thus, $J_{\infty,\epsilon}=-2k\delta^2$, $J_{\bm q}=-2k\delta^2$ for all $\bm q\ne\bm0$ but $J_{\bm0}=0$. Hence, $J_{\infty,\epsilon}=J_{\bm0}-\lim_{\bm q\rightarrow\bm0}J_{\bm q}=2k\delta^2$ and $J_\infty=J^\text{SR}_{n,m}=0$ for all $n, m$. 
Thus, the Ising-Husimi-Temperley coupling constants of our model vanish for the square lattice without angular interactions. {This shows that the existence of cooperativity reflects the inability of the lattice to accommodate the change in geometry associated with a spin state change \cite{HalcrowHysteresis, HalcrowStructure}. When $k_\theta$ is zero the lattice can accommodate a change in size of a molecule with zero energy cost.}

Similarly,  the Ising-Husimi-Temperley interactions  vanish in three or more directions when there are no angular interactions and no interactions beyond nearest neighbor.


The total Hamiltonian for the two-dimensional case is therefore 
\begin{eqnarray}
	H &=& K +  V_{v, \epsilon} + H_\sigma \notag\\
	&=&  K + \frac12\sum_{\mu,\bm{q}}\kappa_{\mu,\bm{q}} v_{\mu,\bm{q}} v_{\mu,-\bm{q}}  
	+ N k \left(\varepsilon_x^2 +  \varepsilon_y^2 \right) \notag\\&&
	+ N\frac{\Delta G}2 \langle\sigma\rangle.
\end{eqnarray}
The only term containing the pseduospins is $H_\sigma$, thus the lowest energy configurations are $\sigma_{i,j}=1$ for all $i,j$ for $\Delta G<0$, $\sigma_{i,j}=-1$ for all $i,j$ for $\Delta G>0$, and all configurations are degenerate for $\Delta G=0$. 
$V_{v, \epsilon}$ is minimized for  $\varepsilon_x = \varepsilon_y = v_{\mu,\bm{q}}$ for all $\mu,\bm{q}$, which yields $V_{v, \epsilon}=0$.
We can immediately read off from the displaced oscillator transformations (Eqs. (\ref{eq:disp-v})-(\ref{eq:disp-eps})) that $V_{v, \epsilon}$ is minimized for 
\begin{eqnarray}
	Q_{\mu,\bm{q}} &=&  -\lim_{\eta\rightarrow 0} \left(\frac{\xi_{\mu}(-\bm{q})}{\kappa_{\mu,\bm{q}} + \eta} \right)\sigma_{\bm{q}}\\
	\epsilon_x  &=& \epsilon_y =
	  \delta \langle\sigma\rangle.
\end{eqnarray}
For the all-HS and all-LS states $\sigma_{\bm{q}}=0$ for $\bm q\ne\bm0$ and $\lim_{\eta\rightarrow 0} {\xi_{\mu}(\bm{0})}/{(\kappa_{\mu,\bm{0}} + \eta)}=0$ so $Q_{\mu,\bm{q}}=0$ for all $\mu,\bm{q}$. Equation (\ref{eq:u-from-Q}) then gives $u^{(\alpha)}_{\nu,i,j}=0$ for all $\alpha,\nu,i$, and $j$.

\section{Parameter values}\label{sec:parametervalues}

In order to extract values for the spring constants, we can use the bulk modulus derived for a cubic lattice with nearest neighbors elastic interactions~\cite{Ye2015} 
\begin{equation}\label{eq:BulkModulus}
     B=\frac{1}{6}k\left(\frac{N}{V_0}\right)^{1/3},
\end{equation}
where $N$ is the total number of unit cells and $V_0$ is the volume of the system at equilibrium. Eq. (\ref{eq:BulkModulus}) differs by a factor of $1/2$ from the expression in \cite{Ye2015}, because in our case we have two springs between neighbouring metal sites. With this expression we can use experimental values to estimate $k$.
For a cubic lattice $V_0/N\approx(2\overline{R})^3$, substituting this into Eq. (\ref{eq:BulkModulus})
\begin{equation}
{k}={12\overline{R}B}.
\end{equation}
Typically bulk moduli in frameworks  and SCO complexes are $B = 2 - 50$~GPa  \cite{Chong,Spencer,Mikolasek,Burtch,Bennett}. As $2\overline{R}$ is approximately the distance between metal sites; typically  $2\overline{R} = 6 - 10$~\AA\xspace. 
Thus, $k= 12-300$~N/m.

For the elastic constant involved in angular interactions, $k_\theta$, we can use the shear modulus of a cubic lattice, see supplementary material  \cite{sup},
\begin{equation}
    {k_\theta}={2\overline{R}G},
\end{equation}
combining with the results reported in Sec. \ref{sec:coupling} this means that $G\neq0$ is required for non-zero Ising interactions. 
Most of the knowledge about the shear modulus of MOFs comes from density functional theory and molecular dynamics computational simulations. For cubic isoreticular MOFs  $G$ was calculated to be $ 6 - 19$~GPa at $T=0$~K, and $6 - 14$~GPa at $T=300$~K \cite{Chong}. For the UiO-66 framework $G$ was calculated to be 18~GPa placing it in the upper limit of the reported values for MOFs \cite{Burtch}. A study on a multicomponent MOF, MUF-32, calculated $G=1.84$~GPa, and decreases to $G=0.54$~GPa upon ligand removal \cite{LeeSeok,Burtch}. Experimentally, the shear modulus of a zeolitic imidazolate framework ZIF-8 was measured to have a minimum of 1~GPa \cite{TanChong}. As such we estimate $G=1 - 19$~GPa, which yields $k_\theta= 1-10$~N/m.  

\begin{figure}
	\includegraphics[width=0.9\columnwidth]{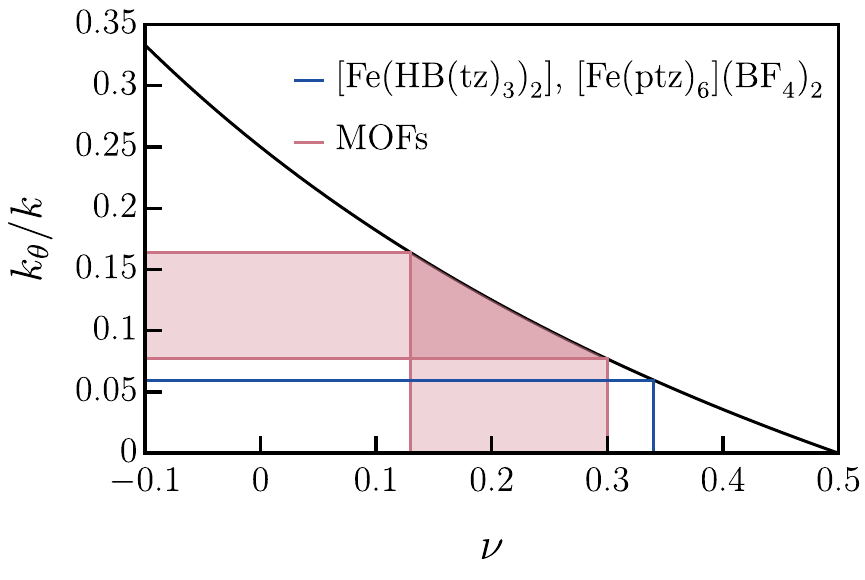}
	\caption{The relationship between Poisson's ratio, $\nu$, and the ratio of spring constants, $k_\theta/k$ in our model, Eq. \ref{eq:konthetaGonB}. The blue line shows the measured value of $\nu$ (and corresponding $k_\theta/k$) for \ch{[Fe(HB(tz)3)2]} \cite{Mikolasek} and \ch{[Fe(ptz)6](BF4)2} \cite{Jung} -- the only SCO materials for which $\nu$ has been reported. The pink shaded region indicates typical values of $\nu$ (and hence $k_\theta/k$) for MOFs. For isotropic, elastic materials the bulk, shear and Young's moduli are positive, which implies that  $-1/10\leq\nu\leq1/2$. This requires that $0\leq k_\theta/k \leq 1/3$.}
	\label{fig:poisson}
\end{figure}

Combining the last two results yields
\begin{equation}\label{eq:konthetaGonB}
    \frac{k_\theta}{k}=\frac{G}{6B}=\frac{1}{4}\frac{(1-2\nu)}{(1+\nu)},
\end{equation}
where $\nu$ is Poisson's ratio \cite{HEARN1997361}, see Fig. \ref{fig:poisson}.
Typically $0.13\lesssim\nu\lesssim0.3$ in MOFs \cite{Mikolasek,Chong}, and $\nu=0.34$ in the two SCO molecular crystals for which we are aware of measurements: \ch{[Fe(HB(tz)3)2]} \cite{Mikolasek} and \ch{[Fe(ptz)6](BF4)2} \cite{Jung}, where tz=1,2,4-triazol-1-yl, and ptz=1-propyltetrazole.
Thus Eq. (\ref{eq:konthetaGonB}) yields ${k_\theta}/{k}\simeq  0.07-0.17$ for frameworks and ${k_\theta}/{k}\simeq 0.06$ for \ch{[Fe(HB(tz)3)2]} and \ch{[Fe(ptz)6](BF4)2}.  Notice that experiments  constrain the ratio ${k_\theta}/{k}$ much more tightly than either spring constant. 

Typically   $2\delta\approx0.2$ \AA\xspace for Fe(II) complexes \cite{COLLET2018, MolnarGabor}
and $2\delta\approx0.1-0.13$ \AA\xspace for  Fe(III) complexes \cite{Harding}.

Using these values we can estimate the magnitude of the Ising coupling constants; here we choose $\overline{R}=49\delta$ and $\nu=0.13-0.34$. This yields
 $J_{10}/\delta^2k = 0.2592$ to $0.1977$, $J_{11}/k\delta^2 = -0.1195$ to $-0.1204$, and $J_{20}/k\delta^2 = 0.1089$ to $0.1333$.
Thus, these Ising constants have similar magnitudes.  Moreover, $J_{n0}>0$ for any $n$, therefore the first few $J_{n0}$ will induce considerable geometrical frustration in the system. Slichter and Drickamer \cite{Slichter,pavlik2013} proposed a simple model for the free energy of an SCO system, where the cooperativity of the system is given by the ``cooperativity parameter" $\Gamma$. Rough estimates of $\Gamma$ for our model, considering only nearest neighbour interactions, are within the same order of magnitude as experimental reported values \cite{Roubeau, linares199}.

\section{Evaluation of the Ising coupling constants}\label{sec:coupling}

For $k_\theta>0$, the Ising coupling constants are non-zero; however, the eigenvalues/eigenvectors of the Hessian cannot be computed analytically. Therefore we perform the integral numerically, unless otherwise stated we use a $2\,000\times2\,000$ Monkhorst-Pack~\cite{Monkhorst} grid.  
(see Sec. \ref{sec:colletive} for further details).

\begin{figure}
\begin{center}
	\includegraphics[width=0.9\columnwidth]{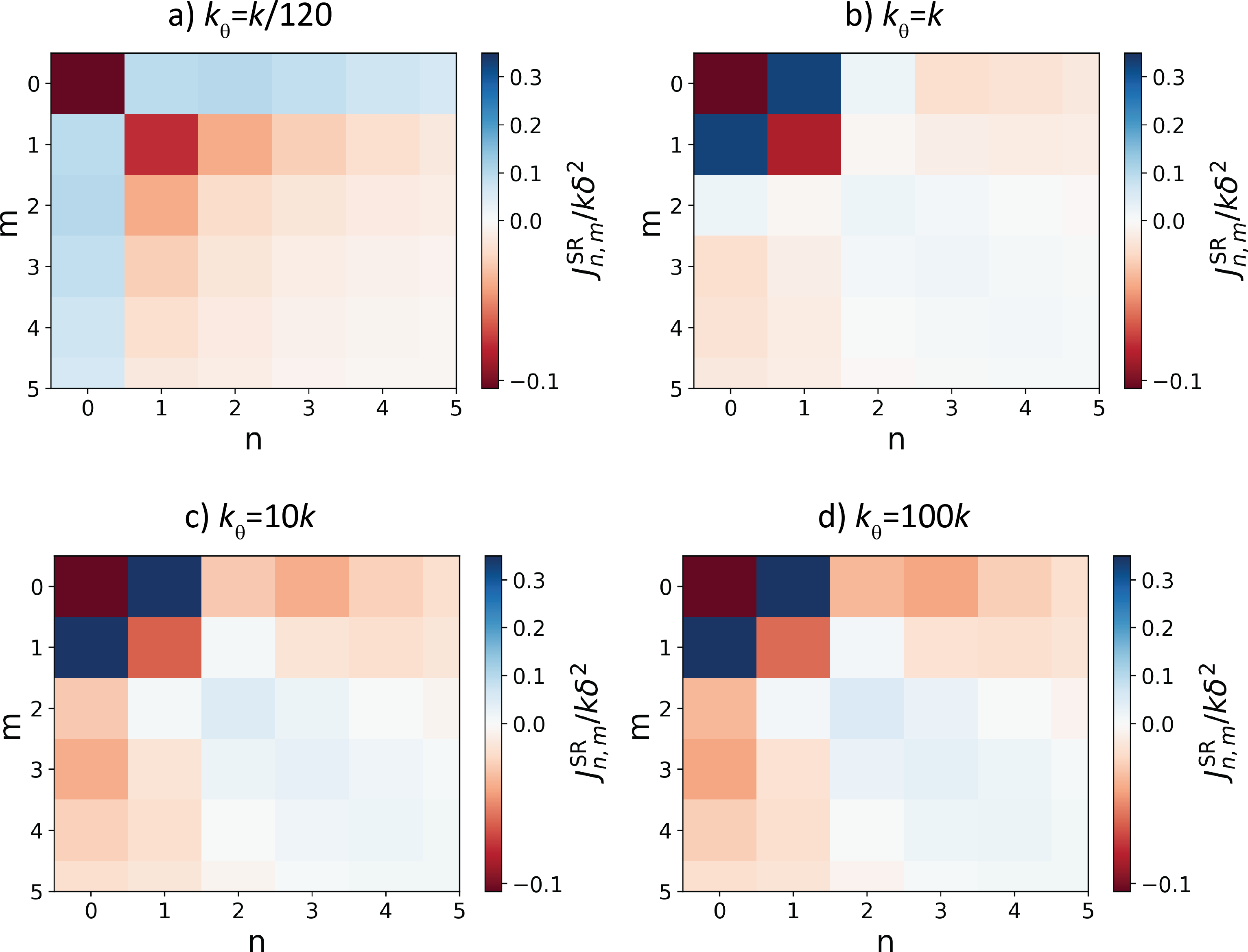} 
	\caption{Short-range spatial dependence of the Ising interactions $J_{n,m}^\text{SR}$, for different values of $k_\theta/k$. There is significant frustration for $k_\theta/k\ll1$, but as $k_\theta/k$ increases this is lifted and the near neighbour interactions cooperate to stabilize checkerboard order.}
	\label{fig:Jnnfordiffkt}
\end{center}
\end{figure}
\begin{figure}
	\begin{center}
		\includegraphics[width=0.9\columnwidth]{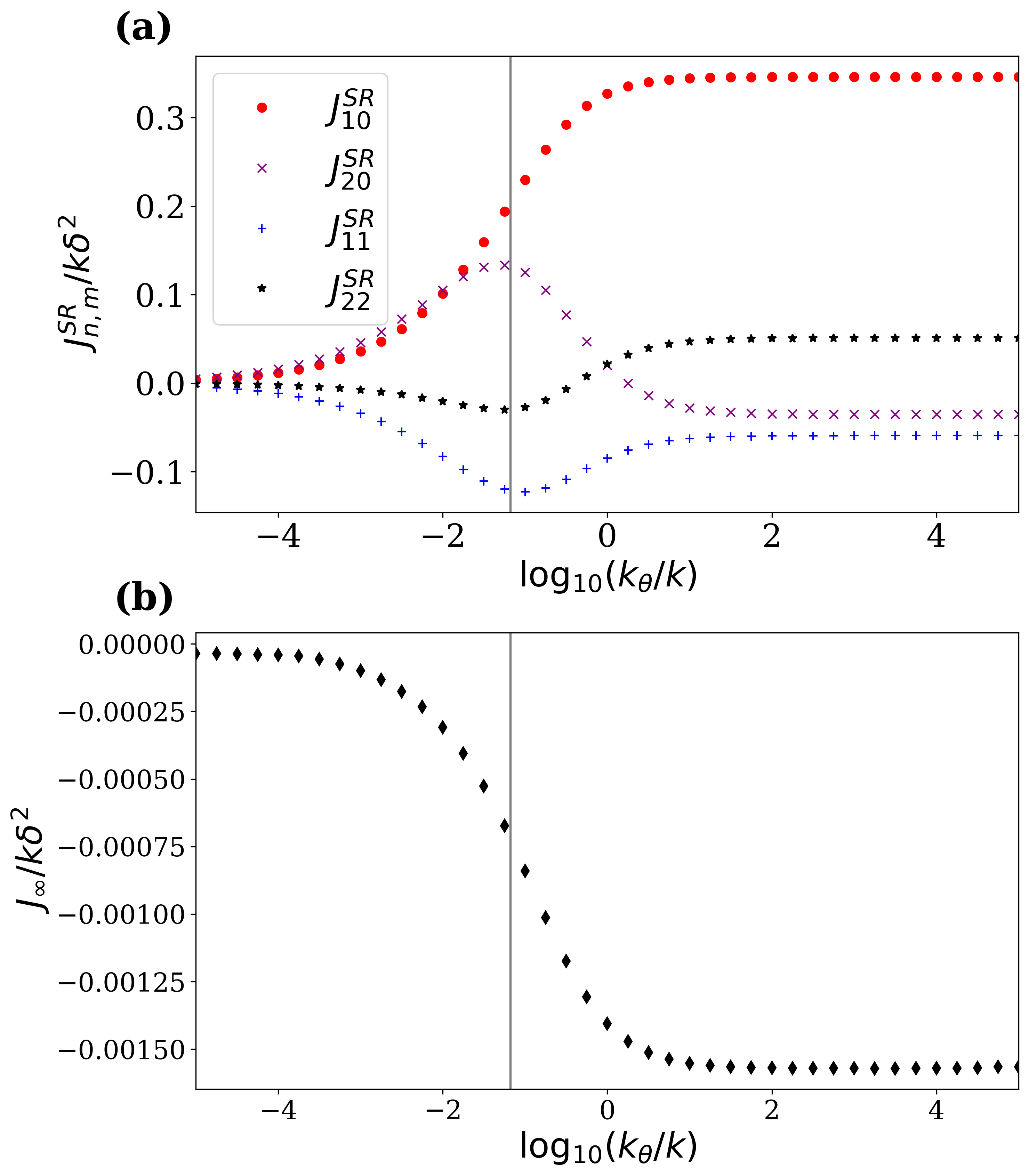} 
		\caption{Dependence of selected Ising couplings  on the relative stiffnesses to  bending, $k_\theta$, and  stretching, $k$. For $k_\theta/k>1$ all the Ising interactions approach constant values. (a)  As $k_\theta$ varies, the signs of all the short ranged Ising constants other than $J_{1,0}^\text{SR}$ and $J_{1,1}^\text{SR}$ change. For increasing $k_\theta$, $J_{1,0}^\text{SR}$ increases, but the other Ising coupling constants vary non-monotonically. 
		For $k_\theta/k<0.013$, $J_{2,0}^\text{SR}$ is slightly larger than $J_{1,0}^\text{SR}$, and as $k_\theta$ approaches $0$, all Ising couplings vanish, consistent with the analytical result for $k_\theta=0$, section \ref{sect:anal}. For $k_\theta/k>1$, the Ising coupling constants approach constant values. $J_{1,0}^\text{SR}$ is always positive and $J_{1,1}^\text{SR}$ is always negative, but the signs of $J_{2,0}^\text{SR}$ and $J_{2,2}^\text{SR}$ depend on $k_\theta/k$. The vertical black line indicates the maximum physically reasonable value for $k_{\theta}/k$ in an isotropic material, see Fig. \ref{fig:poisson}. 
        (b) $J_{\infty}$  is always negative and also  approaches zero as $k_{\theta}$ goes to zero.   
        }
		\label{fig:J10J20J11vskt}
	\end{center}
\end{figure}

We report the couplings for representative $k_\theta/k$ in \cref{fig:Jnnfordiffkt} and the $k_\theta/k$ dependence of selected interactions in \cref{fig:J10J20J11vskt}a.
$J_{1,0}^\text{SR}$ is always positive (i.e., anti-ferroelastic) and $J_{1,1}^\text{SR}$ is always negative (i.e., ferroelastic). However, the signs of many of the other the $J_{n,m}^\text{SR}$ change as $k_\theta/k$ varies. For example, for small $k_\theta/k$, the $J_{n,0}^\text{SR}$ are all positive, but as the ratio approaches unity, they become negative for $n>1$, only $J_{1,0}^\text{SR}$ remains positive. The Husimi-Temperley interaction is always negative and hence favors  a ferroelastic state at all values of $k_{\theta}/k$, \cref{fig:J10J20J11vskt}b. 
When $k_\theta/k$ is sufficiently small, all $J_{n,m}^\text{SR}$ vanish as $k_\theta/k$ approaches zero, consistent with the analytical result for $k_\theta=0$, section \ref{sect:anal}. As $k_\theta/k\rightarrow0$, the shear modulus vanishes and the lattice can accommodate spin state changes with no energy cost. That is, there is no elastic frustration.


Our expression (Eq. (\ref{eq:Jq})) for the Ising-Husimi-Temperley couplings can be decomposed into different contributions from the six different eigenvectors (elastic modes) of the Hessian matrix (Eq. (\ref{eq:hessianelements})). One might hope to gain some physical insight from looking at the relative contributions from the different modes and how they vary as the parameter ratio $k_\theta/k$ varies. Relevant plots of both the different contributions and the mode dispersion relations are shown in the Supplementary Material  \cite{sup}. It is not possible to make many generalisations about the relative contributions and how they vary with $k_\theta/k$. All modes contribute significantly, except when $k_\theta/k$ becomes larger than one, when the contributions of modes with the three largest eigenvalues become negligible. For all the Ising-Husimi-Temperley couplings the relative sign of the contributions from the different modes changes as $k_\theta/k$ varies, indicating a subtle competition between the different modes.

\begin{figure}
	\begin{center}
		\includegraphics[width=0.8\columnwidth]{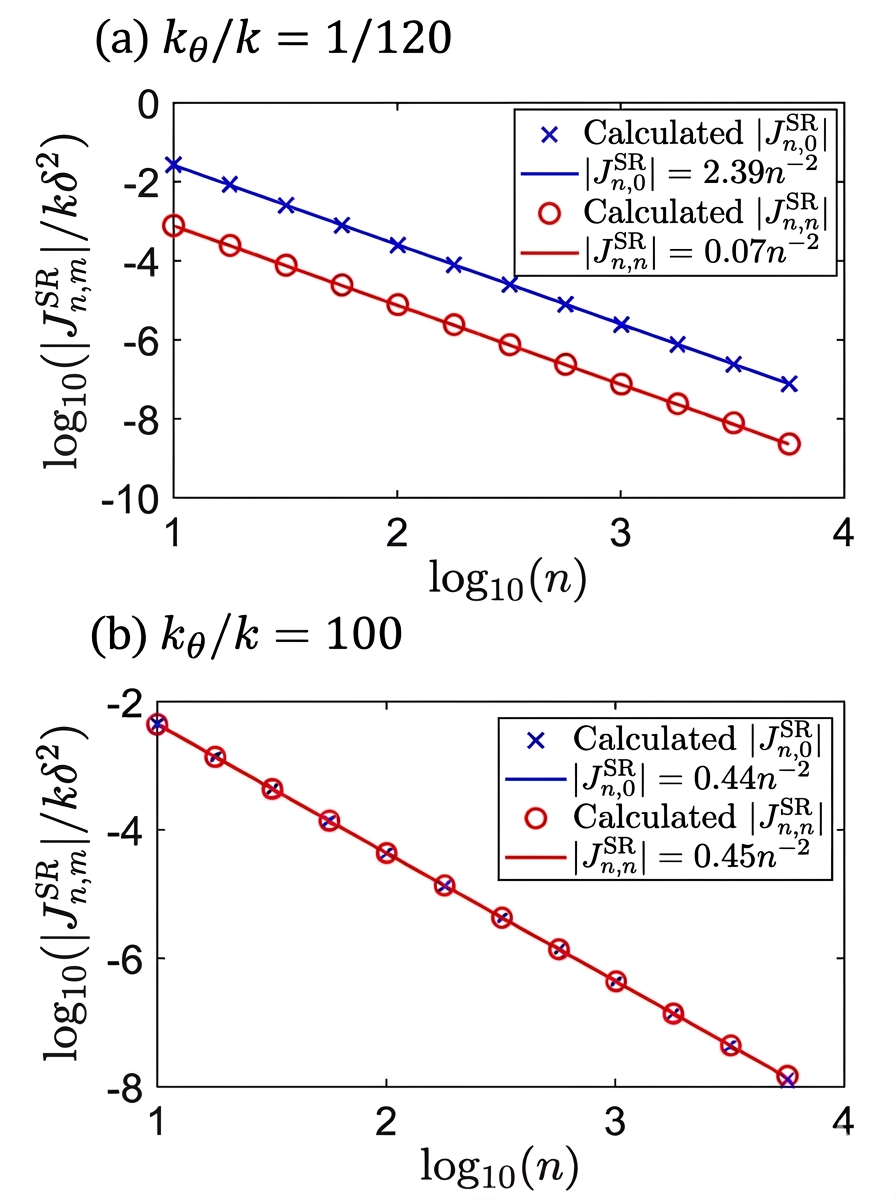} 
		\caption{Logarithmic plots of $|J_{n,0}^\text{SR}|$ and $|J_{n,n}^\text{SR}|$ as a function of $n$, displaying power law behaviour for large $n$, irrespective of the ratio $k_\theta/k$. Prefactors were determined from linear fits performed using a least-squares method. The Ising coupling constants were calculated using a $50\,000\times 50\,000$ grid. }
		\label{fig:LinearFits}
	\end{center}
\end{figure}

At large distances the  Ising coupling constants undergo a power law decay. On the square lattice we find an inverse square law: Fig. \ref{fig:LinearFits} demonstrates that $|J_{n,0}^\text{SR}|= An^{-2}$ and $|J_{n,n}^\text{SR}|= Bn^{-2}$ for large $n$, where $A$ and $B$ are constants that depend on the ratio $k_\theta/k$. 

\section{Collective behavior: spin-state ordering and transitions}\label{sec:colletive}

To study the finite temperature properties of the model we performed Monte Carlo (MC) simulations using the  Metropolis algorithm for different combinations of $k_\theta/k$, $\Delta H$, and $\Delta S$.
The simulations were performed on a  $60\times60$ lattice with periodic boundary conditions.  The Ising coupling constants were calculated from  Eq. \ref{eq:IntegralJnmSR}  using a $2000\times2000$ Monkhorst-Pack grid, and  the cut-off was set to $n_{max} = m_{max} = 29$  to avoid self interaction.  
 

For the cooling-heating simulations we initialized the system at the all-HS phase at high temperatures, where this is the lowest free energy state and we let the system cool. Once the lowest desired temperature is reached we let the system heat.
At each temperature we equilibrate the initial state for $500\times60\times60$ single-spin-flip MC steps.
To take the thermodynamic averages, we run each temperature step for $10\,000\times60\times60$ single-spin-flip MC steps. The next temperature step uses the configuration at the end of the previous temperature. In cases where the system remains HS to the lowest temperatures studied we have also included runs initiated in the  all LS-phase at low temperatures, this can be viewed as a simulation of what one might expect following reverse-light induced spin state trapping (r-LIESST).

\begin{figure*}
	\begin{center}
		\includegraphics[width=6 in]{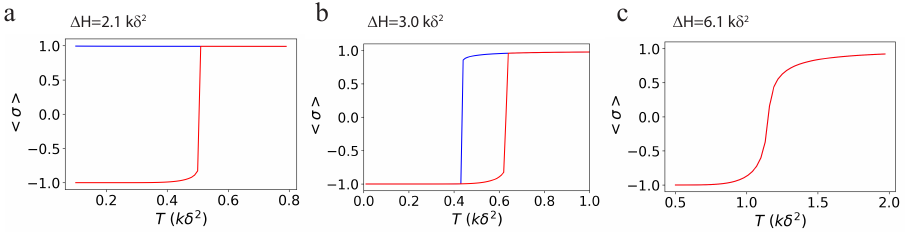}
		\caption{The ensemble average of the HS fraction $\gamma_{HS}=1-2\langle\sigma\rangle$ on thermal heating (red) and cooling (blue)  obtained from Metropolis Monte-Carlo simulations. Here  $k_\theta/k=1/15$ and $\Delta S= \log(200)$.
					}
		\label{fig:simulations1}
	\end{center}
\end{figure*}

The results of these simulations  for different values of  $\Delta H$ are reported in Fig. \ref{fig:simulations1}. 
For reasonable parameters we see three distinct behaviors: thermodynamic crossovers,  Fig. \ref{fig:simulations1}c,  (hysteric) first-order phase transitions, Fig. \ref{fig:simulations1}b, and hidden SCO, Fig. \ref{fig:simulations1}a. 
Crossovers are associated with weakly interacting metal centers, i.e., small $J_{n,m}^\text{SR}/\Delta H$. 
As $\Delta H$ decreases the relative strength of the interactions increases and the crossover is replaced by a first-order phase transition (presumably these are separated by a critical point, but  this is a single point in parameter space and we have not encountered it in our numerics). The width of the hysteresis serves as a rough measure of the relative strength of the interactions between pseudospins. For strong enough interactions the hysteresis becomes so wide that on cooling from high temperatures the system remains in the HS phase -- that is, the lower edge of the hysteresis curve extends below $T=0$. Thus we have hidden SCO,  where the LS state is not accessible through thermal cycling, but can be realized by, for example,  r-LIESST.
It is clear from Fig. \ref{fig:J10J20J11vskt} that all $J_{n,m}^\text{SR}$ become small for small $k_\theta/k$. consistent with this we find crossovers for smaller $\Delta H$ as $k_\theta/k$ decreases (see figures S4 and S5). 

Significantly, the ferroelastic $J_\infty$ term completely changes the phase diagram of the model. In an earlier preprint \cite{v2} we investigated the same model with $J_\infty=0 $ and it exhibited a rich phase diagram with a devil’s staircase structure and many stable spin-state orderings due to the geometric frustration in the Ising interaction. This leads to multi-step transitions in the HS fraction versus temperature curves. However, in the correct Ising model, the ferromagnetic $J_\infty$ term dominates the  Ising interaction and destabilises complex spin-orderings and only uniform high- and low-spin states are stable. However, we note that multistep transitions have been found previously in both Ising-Husimi-Temperley models \cite{Paez2016,JaceKagome,JaceSquare,JacePyro} and direct simulations of elastic models of SCO \cite{Nishino-triangles,NadeemCS,NadeemNano}. Therefore, this suggests that multistep transitions may survive $J_\infty$ in other models.

\section{Isotope effects}\label{sect:Isotope}

{The Ising-Husimi-Temperley coupling constants, $J_{n,m}$, $J_{n,m}^\text{SR}$ and $J_{\infty}$, all scale with $k\delta^2$, which is the elastic energy scale associated with a spin state change of a metal ion. The size of the hysteresis scales with the magnitude of the Ising coupling constants \cite{BOLVIN1995295}, therefore our model predicts that the hysteresis scales with the rigidity of the lattice and with the change in size of the SCO molecules. Thus, a wide hysteresis does not require a crystallographic phase change during spin crossover, consistent with experiments \cite{HalcrowHysteresis}.}

It is important to note that the origin of the Ising coupling constants is purely from the elastic interactions. Therefore they are independent of the mass of the ligands and metals, and do not depend on the dynamics of the system. 

However, isotopic substitutions are expected to change  $\Delta H$ (due to zero-point and thermal vibrational energy changes) and $\Delta S$ (which is dominated by the vibrational contribution to the entropy). This is consistent with the experimental observation (briefly  reviewed  in this Section) that isotopic substitution tends to lead to shifts in crossover temperatures ($T_{1/2}$) and changes in the width of hysteresis, but has not been observed to dramatically change the spin-state ordering.

In the context of collective Jahn-Teller transitions in transition metal spinels, Kanamori \cite{Kanamori}, used displaced operators to transform to new phonon operators. This method yields an expression of the same form as  Eq. (\ref{eq:Jq}) 
\cite{Gehring_1975,Kambara}, where the intermolecular interaction was attributed to the exchange of virtual phonons. In our model this might appear to be also the case, but the Hessian basis shows that the interaction is not dynamical in origin. This has the important implication that isotopic substitution will not change the Ising interactions. 


Isotopic substitution has provided significant insights into molecular and solid state physics. It involves substitution of particular atoms in a compound by the same chemical element with a different nuclear mass (i.e., a nuclear isotope). An example is hydrogen/deuterium substitution which has shown the significant role that quantum nuclear motion
can play in hydrogen bonding, particularly in strong hydrogen bonds (see Ref. \onlinecite{mckenzie2014}
and references therein).
Of particular relevance to the discussion below is that isotopic substitution  not only changes vibrational frequencies but can also change bond lengths.

A key piece of evidence on the road to the BCS theory of superconductivity in 1957 was the observation of an isotope effect.
In 1950 a shift in the transition temperature of mercury was observed, suggesting that superconductivity resulted from electron-phonon interactions, as argued by Fr\"ohlich that same year. In particular, the magnitude of the shift was consistent with theoretical work by Fr\"ohlich.
(Whether he predicted or postdicted the observed effect is a matter of debate, as discussed by Hirsch \cite{Hirsch2011}).
BCS theory gives  $\Delta T_c/T_c = - {1\over 2} \Delta M /M$, where $M$ is the mass of the atom and $\Delta M$ is the mass difference between isotopes. This occurs because phonon frequencies scale as $1/\sqrt{M}$ and is consistent with the mercury experiments.

However, in the 1960s there were many observations of “anomalous” isotope effects, particularly in transition metals, that were inconsistent with the prefactor in this equation. 
These anomalies were resolved by going beyond the BCS theory, and allowing for strong-coupling effects. Following the discovery of cuprate superconductors in 1986, many isotope effects were observed \cite{barbee1988}. However, the consensus now is that these observations do not support an electron-phonon mechanism for superconductivity but rather are due to structural changes due to the isotope substitution. For example, isotopic substitution changes the zero-point energy, and that can alter the unit cell volume and the hopping parameter in a Hubbard model.

This history illustrates the subtleties in interpreting isotope experiments. This is because there are both static and dynamical isotope effects. Changes in isotope can lead to changes in structure, such as bond lengths or lattice constants, and even in changes in crystal symmetry. These structural changes arise because the equilibrium structure of the system is that which minimises the total energy of the system. The contribution to this energy from  the zero-point energy of the
atomic vibrations changes with isotope substitution. Static effects might be classified as those that result from these structural changes. Dynamical effects are those that result from the direct changes in vibrational frequencies.

We now review some of the isotope substitution experiments have been performed on spin-crossover materials \cite{hosoya2003,hosoya2016,molla2021,Kosone,weber2011}.
The first studies are reviewed in section 2.3.5 of Ref. \onlinecite{Gutlich2004}. Isotopic exchange  was investigated for a tris(picolylamine)iron(II) system which exhibits a two-step transition, suggesting anti-ferroelastic Ising interactions. Significant changes in the spin-state transition curve were observed only when the isotopic substitution
(H/D and $^{14}$N/$^{15}$N) was made for atoms directly involved in the hydrogen bonding network that connects the spin-crossover molecules.
For example, with C$_2$H$_5$OD/ND$_2$ the crossover temperature was shifted to higher temperatures by about 15 K and the middle step was no longer present. 

Hosoya \textit{et al.} \cite{hosoya2003} studied the two-dimensional coordination polymer spin-crossover compound Fe(pyridine)$_2$[Ni(CN)$_4$]  and its analogues with the H and N in the pyridine
substituted with D and $^{15}$N, respectively. For both isotope substitutions the spin transition temperature decreased by about 10 K but the hysteresis width changed little. Deuteration decreased the value of  $\Delta H$ and $\Delta S$ (determined from differential scanning calorimetry) by about fifteen and ten percent, respectively. The corresponding decreases for N isotope substitution were about ten and six percent (See Table 1 in Ref. \onlinecite{hosoya2003}). Later, Hosoya \textit{et al.} \cite{hosoya2016} studied a two-dimensional Hofmann-type clathrate host, Fe(4,4’-bipyridine)Ni(CN)$_4$·$n$H$_2$O, with ethanol or acetone guest molecules.
In the absence of guests the complex did not exhibit a spin-state transition. Introduction of ethanol and acetone, led to two-step and one-step spin state transitions, respectively. Deuterium substitutions in the guest molecules changed the transition temperatures and transition widths by amounts of the order of 10 K.

Weber \textit{et al.} \cite{weber2011}, studied the
iron(II) spin-crossover complex [FeL1(HIm)$_2$]  and its isotopic deuterium-labelled analogue [FeL1(DIm)$_2$] where HIm is imidazole. Both exhibit a single-step transition with hysteresis. H/D exchange decreased both the transition temperature and the hysteresis width by a few K. Deuteration decreased the value $\Delta H$ and $\Delta S$ (determined from differential scanning calorimetry) by about twenty and ten percent, respectively. (See Table 2 in Ref. \onlinecite{weber2011}). They estimated an interaction parameter $J = 560$ K, indicating strong intermolecular interactions, which they attributed to  a hydrogen bond involving an oxygen atom of the Schiff base like ligand that serves as a donor for the iron centre. They reference some earlier studies showing how the magnitude of the ligand field in a transition metal complex can be modified by hydrogen bonds involving the complex.

Kosone \textit{et al.} \cite{Kosone} studied a framework material where in the pyridine ligands hydrogen was substituted with deuterium or nitrogen with $^{15}$N. Both substitutions led to significant changes in the shape of the curve describing the temperature dependence of the fraction of high spin molecules. Such a change can occur due to a change in the relative magnitude of $\Delta S$.

Jornet-Moll\'a \textit{et al.} \cite{molla2021} studied the iron(II) salt [Fe(bpp)$_2$](isonicNO)$_2$·HisonicNO·5H$_2$O, which with decreasing temperature undergoes a transition at 162 K. There is a width of about 5 K, associated with hysteresis. With deuteration the transition temperature decreases to 155 K, the width increases to 7 K, and the enthalpy and entropy changes both increase by about fifteen percent.

In terms of an Ising-Husimi-Temperley model picture, the crossover temperature is given by $T_{1/2}= \Delta H/ \Delta S$ and the presence of hysteresis and the difference between the transition temperature for increasing and decreasing temperature sweeps is determined by the  the $J$'s.
Our results imply that there should be no dynamical isotope effects on the $J$’s, i.e. the inter-spin interactions. This does not rule out changes in the crossover temperature $T_{1/2}= \Delta H/ \Delta S$. Both $\Delta H$ and $\Delta S$ will change with isotope substitution as there are significant  vibrational contributions to both. Similarly, our results do not preclude static isotope effects, for example structural changes could cause variation in the spring constants $k$ and $k_\theta$, which would change the Ising interactions.

\section{Conclusions}\label{sec:conclusion}

We have shown that there is an exact mapping, based on the displaced oscillator transformation, between elastic and Ising-Husimi-Temperley models of SCO materials. We have shown this explicitly for a particular model on the  square lattice, but our approach can be  extended to other elastic models and geometries. 

The Ising-Husimi-Temperley model has both short-range and long-range interactions. The degree of geometrical frustration in the  Ising interactions depends strongly on the relative cost of bending and stretching, $k_\theta/k$. For large $k_\theta/k$ nearest neighbour interactions dominate. As $k_\theta/k$ decreases the other near neighbour interactions become important and the system becomes geometrically frustrated.

Indeed, angular interactions are needed in this model to obtain non-zero Ising-Husimi-Temperley couplings. This is because spin-state cooperativity is linked to the inability of the lattice to accommodate the change in geometry associated with a spin state change \cite{HalcrowStructure}. Without angular interactions the lattice can accommodate the geometry change at no energetic cost. That is, the angular interactions introduce elastic frustration into the model. Other interactions that lead to elastic frustration, such as second nearest neighbor spring constants \cite{Ye2015,Paez2016} could also be included related models to similar effect. Note that elastic frustration in the balls and springs model is logically distinct from and does not imply geometrical frustration in the Ising model.
The need for elastic frustration to obtain non-zero Ising couplings  leads to a key structure-property relation: in order to develop complex behaviours the lattice needs to have both strong elastic interactions and be flexible enough to accommodate the molecular distortions.

At large distances the  Ising interaction follows a power law (inverse square law for the square lattice model). However, the Husimi-Temperley interaction  is independent of distance. Nevertheless, it leads to an extensive contribution to the energy as required physically.

The Husimi-Temperley interaction arises from two physical effects. Firstly, the coupling of the average spin-state to the average size of a unit cell. Secondly, a discontinuity at $\bm q=\bm 0$ in the Ising interactions is mediated by changes of the positions of the metals and ligands within the unit cell. In the absence of elastic frustration (e.g., for the one-dimensional chain or for $k_\theta=0$ in higher dimensions) these two contributions are equal and opposite, i.e., there is no net Husimi-Temperley interaction. However, in general the two terms do not cancel.

Both  Ising and Husimi-Temperley interactions arise from the elastic potential alone and not from the dynamical behaviors of the phonons. This means that the Ising interactions are not altered by isotopic substitutions and changes in the  behavior of SCO materials upon isotopic substitution (see Section \ref{sect:Isotope}) are caused by changes in the single ion physics -- specifically $\Delta H$ (due to variations in the vibrational contribution to the enthalpy, particularly zero point energy) and $\Delta S$ (which is dominated by the difference in the vibrational entropy between the two spin-states). 

Although our model is purely classical, it is straightforward to extend the calculation to a quantum treatment of phonons, yielding precisely the same Ising model as the classical calculation. The quantum treatment might tempt one to ascribe the Ising interaction to the exchange of virtual phonons \cite{Kanamori}. This would be incorrect.

The relationship between Poisson's ratio, $\nu$, and $k_\theta/k$ [Eq. (\ref{eq:konthetaGonB})]
allows us to relate the mechanical properties of  materials to their SCO behaviour. As $\nu\rightarrow0.5$ the Ising coupling constants go to zero and only crossovers are observed.
Thus, we predict that controlling  Poisson's ratio, or equivalently the ratio of the  bulk and shear moduli is a key route to engineering emergent phases and multiple anti-ferroelastic phases in SCO materials.

The high porosity and tunable composition of MOFs provide many ways of engineering multi-step transitions. Experimentally, it has been observed that the addition of guests molecules \cite{Zhang, Zhang_2,valverdeMunoz}, the postsynthetic modification of ligands \cite{Clements2016}, and the modification of linkers between SCO centres \cite{Peng, valverdeMunoz} can change the number of steps in SCO transitions and the order observed at intermediate plateaus.
Our work suggests that the key mechanisms by which  these modifications act are modifying the rigidity and ligand field environment of the material.

The explicit model we discuss is relevant to the $[M(L')_n(L)_2]$ and $[M(L')_n\{M'(L)_2\}_2]$ families of SCO frameworks. It is simple enough to be parameterized from standard experiments (measurements of the bulk and shear moduli, heat capacity, and the crystal structure). 
Our model predicts crossovers, first-order transitions that are hysteretic without the need for structural phase transitions, and hidden SCO. However, it does not yield multistep transitions or intermediate phases with anti-ferroelastic orders, which are found in some of these materials. Previous work \cite{JaceSquare} suggests that longer range elastic interactions are vital for stabilizing these phenomena and has demonstrated that they are robust to Husimi-Temperley interactions, like those discussed here. Similarly more exotic phases such as, spin-state ices, have been demonstrated to survive the inclusions of Husimi-Temperley interactions \cite{JaceKagome,JacePyro}.

Applications of our mapping from elastic to Ising models for different geometries and to more detailed elastic models will allow for the study of the effect that structure has on SCO transitions, and how the strength of the interactions and the volume change of the SCO molecules influence hysteresis and spin ordering. 

Finally, our model is not only relevant for SCO systems, as it could be applied to any binary system with lattice mismatch. For example, elastic models have been shown to be relevant to the understanding of heterogeneous solids (e.g., binary alloys) \cite{FRATZL,Frechette1}. Thus, our work suggests that changes to the lattice constants with composition will  be more important for such materials than has previously been appreciated.

\begin{acknowledgments}

This work was supported by the Australian Research Council through grants number DP200100305 and DP230100139. We thank Nena Batenburg, Cameron Kepert, Suzanne Neville, Lachlan Parker, and Jesse Woods for helpful discussions and  Kamel Boukheddaden, Tazmin Bradbury, Zejun Huang, Roderich Moessner, Dvira Segal, Rajiv Singh, and William Vivian for helpful comments on a draft of this manuscript. Parts of the derivation reported in Appendix \ref{sect:lr} were first performed with Google Gemini. Each step was then checked by
a human.

\end{acknowledgments}

\appendix
\section{Long distance behavior of Ising interaction}\label{sect:lr}

The notation  in this appendix is greatly simplified if we  promote the lattice separation into a continuous variable, $\bm{n}_{n,m}\rightarrow\bm R$; and similarly  $J^\text{SR}_{n,m} \rightarrow J^\text{SR}(\bm{R})$. Clearly, only the values of $J^\text{SR}(\bm{n}_{n,m})=J^\text{SR}_{n,m}$ are of physical significance.

Consider the Ising interaction $J^\text{SR}_{\bm{q}}$ as a function of polar co-ordinates, 
$\bm{q}=(q \cos \phi_q, q \sin \phi_q)$.
Our numerical results (see, e.g., Figs \ref{fig:Jq} and \ref{fig: JqPhi}) show that 
it is approximately constant as a function of $q$ and 
the angular dependence has a component that is independent of $\phi_q$  and a component with periodicity $\pi/2$.
Here we show that this implies that, in real space, $J^\text{SR}(\bm{R})\sim 1/|\bm{R}|^2$ or, equivalently, $J^\text{SR}_{n,m} \sim 1/|\bm{n}_{nm}|^2$ for large distances $|\bm{R}|$ or $|\bm{n}_{n,m}| = \sqrt{n^2+m^2} \gg 1 $.

The expression for $J^\text{SR}_{n,m} \rightarrow J^\text{SR}(\bm{R})$, given in Equation (\ref{eq:IntegralJnmSR})  in terms of the Fourier transform of $J^\text{SR}_{\bm{q}}$
involves an integral over the first Brillioun zone, which in this case is a square with sides of length $2\pi$. 
Introducing the polar co-ordinates, 
$\bm{R}=(R \cos \phi, R \sin \phi)$,
the Fourier transform can be written
\begin{multline}
	J^\text{SR}(R, \phi) =  \frac{1}{4 \pi^2}\int_{0}^{q_c(\phi_q)}
	\int_{0}^{2\pi} J^\text{SR}_{(q,\phi_q)} \\ \times e^{-i q R \cos(\phi_q - \phi)} q \, d\phi_q \, dq,
\end{multline}
where the upper limit of the $q$ integral is
\begin{equation}
	q_c(\phi_q) \equiv \pi/{\rm{max}(|\cos \phi_q |, |\sin \phi_q |}).
\end{equation}
Due to the weak dependence on $q$ we write
$J^\text{SR}(q, \phi_q) \simeq g(\phi_q)$
and make the change of variables $x = qR$, 
\begin{multline}
	J^\text{SR}(R, \phi) \simeq  \frac{1}{4 \pi^2 R^2}
	\int_{0}^{q_c(\phi_q)R}
	\int_{0}^{2\pi} g(\phi_q) \\ \times e^{-i x \cos(\phi_q - \phi)} x \, d\phi_q \, dx.
\end{multline}
When $R$ is large the upper bound on the $x$ integral tends to infinity, leading to
\begin{equation}
	J^\text{SR}(R, \phi) \simeq \frac{1}{4 \pi^2 R^2}
	\int_{0}^{\infty}
	\int_{0}^{2\pi} g(\phi_q) e^{-i x \cos(\phi_q - \phi)} x \, d\phi_q \, dx.
	\label{transform}
\end{equation}

We now show how the angular dependence of the transform is dictated by the angular Fourier components of $g(\phi_q)$.


Because $g(\phi_q)$ is $2\pi$-periodic, it can be decomposed into an angular Fourier series:
\begin{equation}
	g(\phi_q) = 
	\sum_{\nu=-\infty}^{\infty} a_\nu e^{i\nu\phi_q}.
\end{equation}

For a system with four-fold rotational symmetry and $g(\phi_q)$ a real function
\begin{equation}
	g(\phi_q) = a_0 +
	2 \sum_{\mu=1}^{\infty}  a_{4\mu} \cos (4 \mu \phi_q).
	\label{eq:4fold}
\end{equation}
We show below that if Eqs. (\ref{transform}) and (\ref{eq:4fold}) hold exactly, then
\begin{equation}
	J^\text{SR}(R,\phi) = \frac{4}{ \pi R^2}  
	\sum_{\mu=1}^{\infty}  \mu \, a_{4\mu} \cos (4 \mu \phi).
	\label{transform3}
\end{equation}


\subsection{Analysis}
Numerically, we find that $a_0 <0$ (see, e.g., Figs. \ref{fig:Jq} and \ref{fig: JqPhi}). If we assume that the dominant
Fourier component is $a_4$ then it is positive.

This means that the coefficient of $1/R^2$ in $J^\text{SR}(R,\phi$) for 
$\phi=0$ is positive and for $\phi=\pi/2$ it is negative.
In other words, the interactions along the Cartesian axes, $J_{n,0}$, are ``antiferromagnetic" but the interactions along the diagonals, $J_{n,n}$, are ``ferromagnetic".

Furthermore, the magnitude of the $a_4$ term determines the difference
in magnitude between $J_{n,0}$ and $J_{n,n}$ (Fig. \ref{fig:LinearFits}).

\subsection{Derivation}

Substituting the angular series into the integral \eqref{transform} yields:
\begin{multline}
	J^\text{SR}(R, \phi) \simeq
	\frac{1}{4 \pi^2 R^2}
	\sum_{\nu=-\infty}^{\infty} a_\nu \\ \times \int_{0}^{\infty} x \left[ \int_{0}^{2\pi} e^{i\nu\phi_q} e^{-i x \cos(\phi_q - \phi)} \, d\phi_q \right] dx .
	\label{transform5}
\end{multline}

Using the Jacobi-Anger identity, \cite[Eq.~(9.1.41)]{AbramowitzStegun1964},
\begin{equation}
	\exp(i x \cos \alpha) = \sum_{n=-\infty}^{\infty} i^n J_n(x) e^{in\alpha},
\end{equation}
the inner angular integral evaluates to a Bessel function of the first kind $J_n$:
$$\int_{0}^{2\pi} e^{i\nu\phi_q} e^{-i x \cos(\phi_q - \phi)} \, d\phi_q = 2\pi (-i)^\nu e^{i\nu\phi} J_\nu(x)$$
Substituting this back into \eqref{transform5} gives a radial integral for each angular mode $\nu$:
\begin{equation}
	J^\text{SR}(R, \phi) \simeq \frac{1}{2 \pi R^2}  \sum_{\nu=-\infty}^{\infty} a_\nu (-i)^\nu e^{i\nu\phi} \int_{0}^{\infty} x  J_\nu(x) \, dx.
	\label{transform2}
\end{equation}

The remaining integral,
\begin{equation}
	b_\nu \equiv \int_{0}^{\infty} x J_\nu(x) \, dx,
\end{equation}
is a purely numerical constant. 
In the standard Riemann or Lebesgue sense, this integral does not converge because the integrand oscillates with an amplitude that grows infinitely large as $x \to \infty$.
To evaluate the integral rigorously, we introduce a regularizing exponential damping factor $e^{-\eta x}$ where $\eta > 0$, and then take the limit as $\eta \to 0^+$. 
\cite{GelfandShilov1964}. 
The resulting expression corresponds to a Laplace transform that can be evaluated via standard identities \cite{BatemanTransforms1954}.

We first evaluate the $\nu=0$ integral, showing that it vanishes.
\begin{equation}
	b_0 = \lim_{\eta \to 0^+} \int_{0}^{\infty} x J_0(x) e^{-\eta x} \, dx.
\end{equation}
Using the standard Laplace transform identity for Bessel functions, $\int_{0}^{\infty} x J_0(x) e^{-\eta x} \, dx = \frac{\eta}{(\eta^2 + 1)^{3/2}}$.
Taking the limit:
\begin{equation}
	b_0 = \lim_{\eta \to 0^+} \frac{\eta}{(\eta^2 + 1)^{3/2}}  
= 0.
\end{equation}

This has to be the case because if all the $a_\nu$ are zero except for $\nu=0$, then $J^\text{SR}_{\bm{q}}=g(\phi_q)$ is a constant
and the Fourier transform of a constant function is a delta function, i.e., there are no long-range terms. 

For any non-negative integer $\nu$, the regularized value of this family of integrals follows the remarkably simple identity:
\begin{equation}
	b_\nu=\lim_{\eta \to 0^+} \int_{0}^{\infty} x J_\nu(x) e^{-\eta x} \, dx = \nu.
\end{equation}

The integral inside the limit can be recognized as the Laplace transform of the function $f(x) = x J_n(x)$, where $\eta$ acts as the transform variable. 
\begin{equation}
	\int_{0}^{\infty} e^{-\eta x} \left[ x J_n(x) \right] dx = \mathcal{L}\left\{ x J_n(x) \right\}(\eta).
\end{equation}
Using the Laplace transform property for multiplication by $x$:
\begin{equation}
	\mathcal{L}\left\{ x J_\nu(x) \right\} = -\frac{d}{d\eta} \mathcal{L}\left\{ J_\nu(x) \right\}.
\end{equation}
The Laplace transform of the Bessel function of the first kind $J_\nu(x)$ for $ \nu \geq 0$ is:
\begin{equation}
	\mathcal{L}\left\{ J_\nu(x) \right\} = \frac{\left(\sqrt{\eta^2 + 1} - \eta\right)^\nu}{\sqrt{\eta^2 + 1}}.
\end{equation}
Thus, 
\begin{equation}
	\int_{0}^{\infty} x J_\nu(x) e^{-\eta x} \, dx = -\frac{d}{d\eta} \left[ \frac{\left(\sqrt{\eta^2 + 1} - \eta\right)^\nu}{\sqrt{\eta^2 + 1}} \right].
\end{equation}
To differentiate this expression with respect to $\eta$ using the quotient rule, we let $u = \sqrt{\eta^2 + 1} - \eta$ and note that $\frac{du}{d\eta} = \frac{\eta}{\sqrt{\eta^2 + 1}} - 1 = -\frac{u}{\sqrt{\eta^2 + 1}}$.
Evaluating the complete derivative yields:
\begin{multline}
	\int_{0}^{\infty} x J_\nu(x) e^{-\eta x} \, dx \\= \frac{\nu \left(\sqrt{\eta^2 + 1} - \eta\right)^\nu}{\eta^2 + 1} + \frac{\eta \left(\sqrt{\eta^2 + 1} - \eta\right)^\nu}{(\eta^2 + 1)^{3/2}}.
\end{multline}

Now, take the limit as $\eta \to 0^+$:
\begin{equation}
	b_\nu=\lim_{\eta \to 0^+} \left[ \frac{\nu \left(\sqrt{\eta^2 + 1} - \eta\right)^\nu}{\eta^2 + 1} + \frac{\eta \left(\sqrt{\eta^2 + 1} - \eta\right)^\nu}{(\eta^2 + 1)^{3/2}} \right] =\nu.
\end{equation}

For negative integer $\nu$, $J_{-\nu} = (-1)^\nu J_\nu(z)$, and so $b_\nu =  (-1)^\nu \nu$.
Hence, 
\eqref{transform2}
becomes
\begin{equation}
	J^\text{SR}(R, \phi) \simeq \frac{1}{2 \pi  R^2} \sum_{\nu=1}^{\infty} \nu \left[ a_\nu (-i)^\nu e^{i\nu\phi} +  a_{-\nu} i^{\nu} e^{-i\nu\phi} \right].
	\label{transform4}
\end{equation}
For the case of four-fold rotational symmetry this reduces to \eqref{transform3}. 

\bibliography{GianRefs}

%
%
%
\pagebreak
\onecolumngrid
\begin{center}
	\textbf{\large Supplemental Materials: Exact mapping from short-ranged harmonic elastic models of spin crossover materials to Ising models with  interactions at all length scales}
\end{center}
\twocolumngrid

\renewcommand\thefigure{S\arabic{figure}}
\renewcommand\thetable{S\arabic{table}}
\renewcommand\theequation{S\arabic{equation}}
\renewcommand\thesection{S\arabic{section}}
\setcounter{figure}{0}
\setcounter{table}{0}
\setcounter{equation}{0}
\setcounter{section}{0}
\renewcommand{\appendixname}{}

\section{Shear modulus}\label{sec:ShearModulus}

\begin{figure}[b]
	\centering
	\includegraphics[width=0.9\columnwidth]{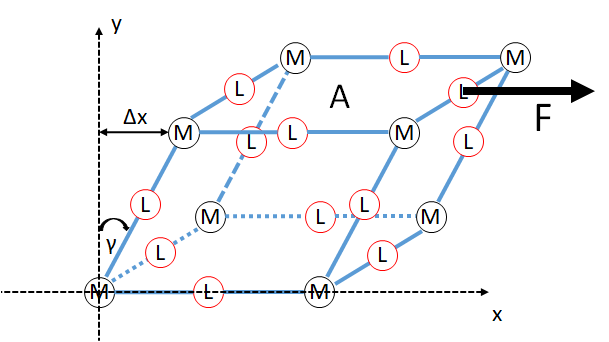}
	\caption{Shear strain on a unit cell of a cubic lattice with the arrangement of ligands (L) and metals (M) described in the paper. $A$ is the area a single layer, $F$ is an externally applied force, $\Delta x$ is the displacement of the layer from the $y$ axis, $\gamma$ is the angular displacement due to force $F$.}
	\label{ShearStrain}
\end{figure}

The shear modulus is defined as
\begin{equation}
	G=\frac{F/A}{\Delta x/l}
\end{equation}
where $F$ is a force applied on a surface $A$ leading to an angular displacement $\gamma$ such that $\tan\gamma=\Delta x/l$ as indicated in Fig. \ref{ShearStrain}. For small $\gamma$ we can make the approximation $\tan\gamma\approx\gamma$, and we let $r_o=\overline{R}+\delta\langle\sigma\rangle$ be the distance between a metal and ligand, therefore
\begin{equation}
	G=\frac{F}{\gamma A}
\end{equation}

\begin{figure}[b]
	\centering
	\includegraphics[width=0.4\textwidth]{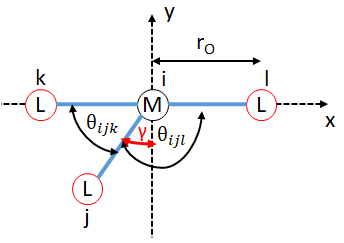}
	\caption{Schematic of a single metal site in the top layer of a cubic lattice.}
	\label{ShearOneSite}
\end{figure}

The force $F$ at mechanical equilibrium is equal in magnitude and opposite in direction to the $x$ component of the force on a single layer from all other particles in the system. We assume that none of the springs are either stretched or compressed, such that there are no contributions to the force due to stretching. Now consider a metal site and its six nearest ligands, a strain  will lead to an angle dependent force on metal $M$, but the only particles inlvolved are those in the $x-y$ plane, see Fig. \ref{ShearStrain}, because the angle between the other bonds is $\pi/2$. 
From our Hamiltonian [Eq. (1)], the energy contribution due to angle $\theta_{ijk}$ between the bonds formed by $M_i-L_j$ and $M_i-L_k$, see Fig. \ref{ShearOneSite}, is given by
\begin{equation}\label{AngleEnergy}
	V(\theta_{ijk})=\Bar{R}^2\frac{k_\theta}{2}\sin^2\left(\theta_{ijk}-\frac{\pi}{2}\right).
\end{equation}

In general
\begin{equation}
	\cos\theta_{ijk}=\frac{(\bm{r}_k-\bm{r}_i)\cdot(\bm{r}_j-\bm{r}_i)}{|\bm{r}_k-\bm{r}_i||\bm{r}_j-\bm{r}_i|}=\frac{(\bm{r}_k-\bm{r}_i)\cdot(\bm{r}_j-\bm{r}_i)}{r_o^2},
\end{equation}
where $r_o$ is the equilibrium spacing between a metal and the ligands it is covalently bound to,
and, for small $\gamma$,
\begin{equation}
	\cos\theta_{ijk}=\cos\left(\frac{\pi}{2}-\gamma\right)=\sin\gamma\approx\gamma=\frac{\pi}{2}-\theta_{ijk}.
\end{equation}
Therefore,
\begin{equation}
	\frac{\partial \theta_{ijk}}{\partial \bm{r_i}}
	=\frac{\bm{r}_{ji} + \bm{r}_{ki}}{r_o^2}
	=-\frac{\hat{\bm x}(1+\sin\gamma)+\hat{\bm y}\cos\gamma}{r_o}
\end{equation}
and the force on metal $i$ due to ligands $j$ and $k$ is 
\begin{equation}
	\begin{split}
		\bm{f}_{i;jk}&=-\frac{\partial V}{\partial \theta_{ijk}}\frac{\partial \theta_{ijk}}{\partial \bm{r_i}}\\
		&=-\frac{\bar{R}^2k_\theta\gamma(1+\sin\gamma)}{r_o}\hat{\bm x}
		-\frac{\bar{R}^2k_\theta\gamma\cos\gamma}{r_o}\hat{\bm y}
	\end{split}
\end{equation}

Following the same procedure for the angle $\theta_{ijl}$, see Fig. \ref{ShearOneSite}, yields
\begin{equation}
	\bm{f}_{i;jl}=-\frac{\Bar{R}^2k_\theta\gamma(1-\sin\gamma)}{r_o}\hat{\bm x}+\frac{\Bar{R}^2k_\theta\gamma\cos\gamma}{r_o}\hat{\bm y}.
\end{equation}
Hence the total force on metal $i$ due to displacement $\gamma$ is
\begin{equation}
	\bm{f}_i=-\frac{2\bar{R}^2k_\theta\gamma}{r_o}\hat{\bm x}.
\end{equation}

A single layer of a cubic system has $N^{2/3}$ metals, where $N$ is the total number of metals. Therefore, the force applied on the system is $\bm{F}=-N^{2/3}\bm{f}_i$.
The surface area of a single layer is $A=N^{2/3}4r_o^2$ therefore 
$G=
{\Bar{R}^2k_\theta}/{2r_o^3}$.
For $\delta\ll\overline{R}$ we can approximate $r_o=\overline{R}+\delta\langle\sigma\rangle\approx\overline{R}$, and hence
\begin{equation}
	G=\frac{k_\theta}{2\overline{R}}.
\end{equation}

If the shear modulus has not been directly measured it can be calculated from Young's modulus $Y$, bulk modulus $B$, and Poisson's ratio $\nu$  \cite{Landau} via
\begin{equation}
	G=\frac{3BY}{9B-Y},
\end{equation}
\begin{equation}
	G=\frac{Y}{2(1+\nu)}.
\end{equation}

\begin{figure*}
	\centering
	\includegraphics[width=0.9\textwidth]{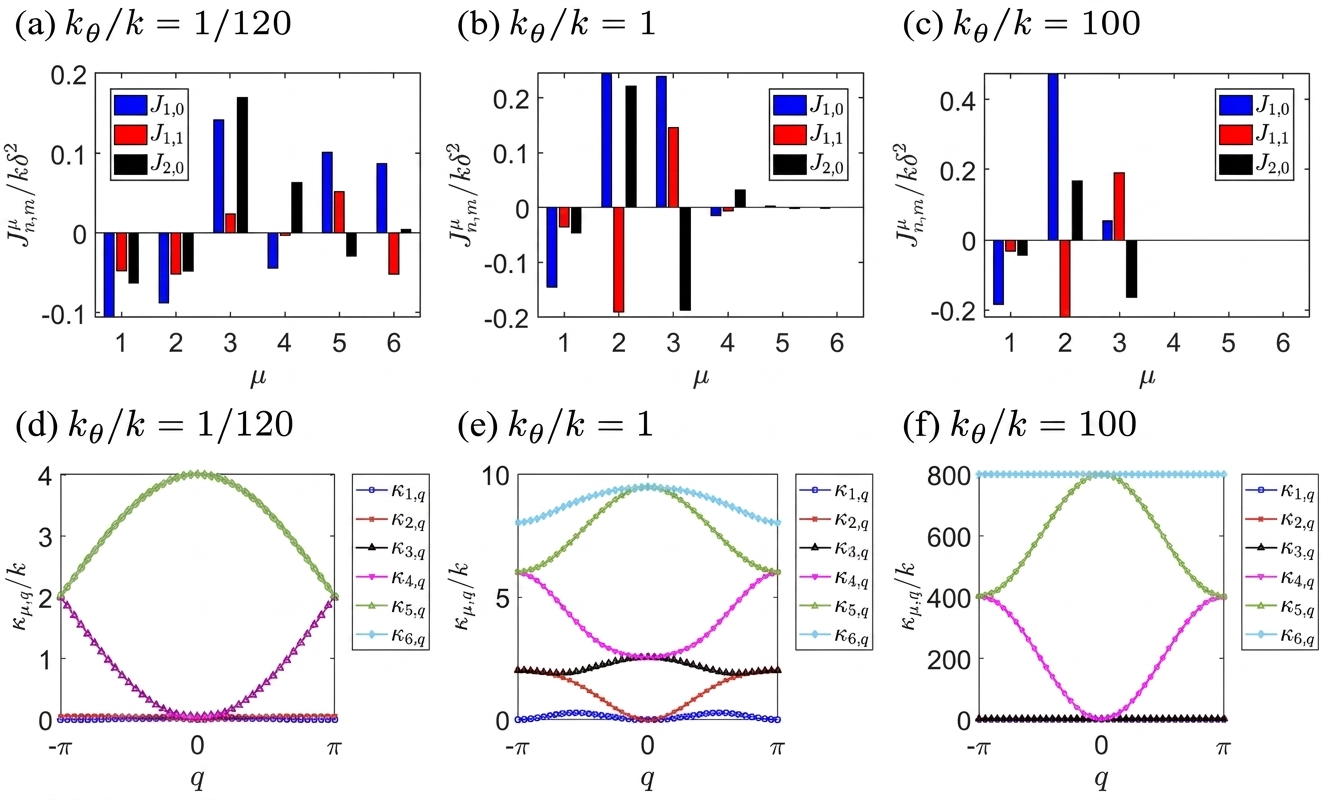}
	\caption{(a-c) The contributions to $J_{1,0}$, $J_{1,1}$, and $J_{2,0}$ from the different eigevalues/eigenvectors of the Hessian matrix, [Eq. (28)], are shown for different $k_\theta/k$. (d-f) Eigenvalues of the Hessian matrix, $\kappa_{\mu,q}$, as a function of $q$ along $q=q_x=q_y$, for different values of $k_\theta/k$. For large $k_\theta/k$, the dominant contributions come from $\mu=1,2,3$; this is because of the large magnitudes of eigenvalues $\mu=3, 4, 5$, as shown in the bottom plots; and because $J^{\mu}_{n,m}$ is inversely proportional to the eigenvalues of the Hessian matrix, [Eq. (29)]. Here, we can also see the subtle competition between the different contributions, and how their relative signs stop changing for large $k_\theta$. }
	\label{fig:ContributionPlots}
\end{figure*}

\section{Decomposition of the Ising-Husimi-Temperley coupling constants}\label{sec:decomposition}
The contributions from the different $J^{\mu}_{n,m}$ change as the ratio $k_\theta/k$ changes,  Figs. \ref{fig:ContributionPlots}a to \ref{fig:ContributionPlots}c. The contributions from $\mu=4,5,6$ decrease as $k_\theta/k$ increases. This is because $J^{\mu}_{n,m}$ is inversely proportional to $\kappa_{\mu,q}$, [Eq. (29)], and, for large $k_\theta/k$, the magnitudes of $\kappa_{\mu,q}$ for $\mu=4,5,6$ are much larger than for $\mu=1,2,3$,  Figs. \ref{fig:ContributionPlots}d to \ref{fig:ContributionPlots}f. There is also a subtle competition between the different $J^{\mu}_{n,m}$, as their signs alternate and their magnitudes differ from each other. For the three cases considered, the magnitudes and signs of $J^{\mu}_{n,m}$ change drastically as $k_\theta/k$ varies, where some $J^{\mu}_{n,m}$ can become considerably larger in magnitude than the others, but their signs alternate which makes the resultant Ising coupling constants smaller.

%

\begin{figure*}
	\begin{center}
		\includegraphics[width=6 in]{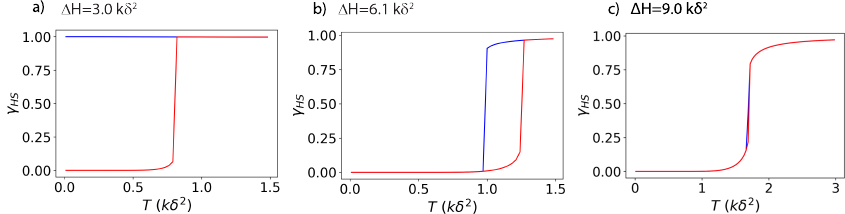}  
		\caption{The ensemble average of the HS fraction $\gamma_{HS}=1-2\langle\sigma\rangle$ on thermal heating (red) and cooling (blue)  obtained from Metropolis Monte-Carlo simulations. Here  $k_\theta/k=1$ and $\Delta S= \log(200)$.
		}
		\label{fig:simulations Ktheta1}
	\end{center}
\end{figure*}

\begin{figure*}
	\begin{center}
		\includegraphics[width=6 in]{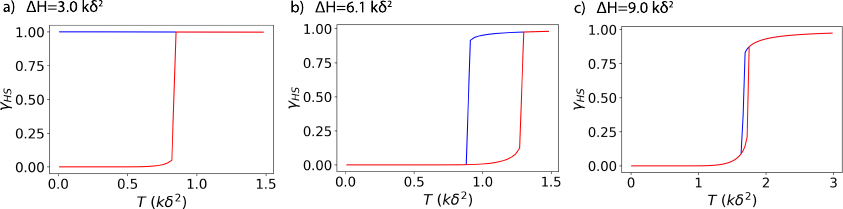}  
		\caption{The ensemble average of the HS fraction $\gamma_{HS}=1-2\langle\sigma\rangle$ on thermal heating (red) and cooling (blue)  obtained from Metropolis Monte-Carlo simulations. Here  $k_\theta/k=100$ and $\Delta S= \log(200)$.
		}
		\label{fig:simulations Ktheta100}
	\end{center}
\end{figure*}

\end{document}